\documentclass[12pt]{article}
\def\spacingset#1{\renewcommand{\baselinestretch}%
{#1}\small\normalsize} \spacingset{1}

\newcommand{\blind}{1}

\usepackage{mathrsfs}
\usepackage{amssymb,amsmath,amsthm}
\usepackage{multirow,multicol}
\usepackage{natbib}
\usepackage{placeins}
\usepackage{rotating}
\usepackage[center]{subfigure}
\usepackage{amsfonts}

\newtheorem{theorem}{\textbf{Theorem}}

\newtheorem{example}{\textbf{Example}}

\newtheorem{assumption}{\textbf{Assumption}}

\newtheorem{step}{\textbf{Step}}

\def\logit{{\rm logit}}

\usepackage{etoolbox}
\makeatletter
\patchcmd{\maketitle}{\@fnsymbol}{\@alph}{}{}
\makeatother

\begin{document}

\if1\blind
{
  \title{\bf Estimation of Optimal
Individualized Treatment Rules Using a Covariate-Specific Treatment Effect Curve with High-dimensional Covariates}
  \author{Wenchuan Guo\\
    Department of Statistics, University of California Riverside,\\
and Global Biometric Sciences, Bristol-Myers Squibb,\\
    Xiao-Hua Zhou   \\
    Beijing International Center for Mathematical Research, \\
and Department of Biostatistics, Peking University,\\
    Shujie Ma\\
    Department of Statistics, University of California Riverside\\
    \noindent
 \thanks{
    Correspondence should be addressed to Xiao-Hua Zhou and Shujie Ma.  Wenchuan Guo is Senior Biostatistician, Bristol-Myers Squibb, Pennington, NJ 08534. Xiao-Hua Zhou is PKU Endowed Chair Professor, Peking University. (Email: azhou@math.pku.edu.cn). Shujie Ma is Associate Professor,
Department of Statistics, University of California-Riverside, Riverside, CA
92521. (Email: shujie.ma@ucr.edu).  Guo and Ma's research was supported by NSF grants DMS-1712558 and and DMS-2014221, NIH grant R01 ES024732-03 and UCR Academic Senate CoR grant. Zhou's research was supported by NSFC 81773546. The authors are grateful to the Editor,
the Associate Editor, and two anonymous reviewers for their constructive
comments that helped us improve the article substantially.}\hspace{.2cm} }
  \maketitle
} \fi
\if0\blind
{
  \title{\bf Estimation of Optimal
Individualized Treatment Rules Using a Covariate-Specific Treatment Effect Curve with High-dimensional Covariates}
 
  \maketitle
} \fi

\clearpage
\bigskip
\begin{abstract}
With a large number of baseline covariates, we propose a new semi-parametric modeling strategy for heterogeneous treatment effect estimation and individualized treatment selection, which are two major goals in personalized medicine. We achieve the first goal through estimating a covariate-specific treatment effect (CSTE) curve modeled as an unknown function of a weighted linear combination of all baseline covariates. The weight or the coefficient for each covariate is estimated by fitting a sparse semi-parametric logistic single-index coefficient model. The CSTE curve is estimated by a spline-backfitted kernel procedure, which enables us to further construct a simultaneous confidence band (SCB) for the CSTE curve under a desired confidence level. Based on the SCB, we find the subgroups of patients that benefit from each treatment, so that we can make individualized treatment selection.  The innovations of the proposed method are three-fold.  First, the proposed method can quantify variability associated with the estimated optimal individualized treatment rule with high-dimensional covariates. Second, the proposed method is very flexible to depict both local and global associations between the treatment and baseline covariates in the presence of high-dimensional covariates, and thus it enjoys flexibility while achieving dimensionality reduction. Third, the SCB achieves the nominal confidence level asymptotically, and it provides a uniform inferential tool in making individualized treatment decisions.
\end{abstract}

\noindent%
\vfill

\newpage
\spacingset{1.5} 
\section{Introduction\label{SEC:Intro}}
Personalized medicine aims to tailor medical treatments according to patient characteristics, and it has gained much attention in modern biomedical research. The success of personalized medicine crucially depends on the development of reliable statistical tools for estimating an optimal treatment
regime given the data collected from clinical trials or observational studies \citep{KL19}. In the literature, there are two general statistical approaches and their hybrids for deriving an optimal Individualized Treatment Rule (ITR) based on clinical trial data. The first general approach targets at direct optimization of the population average outcome under an ITR.  Several independent research groups have  proposed a new framework of deriving the ITR that recasts the problem of maximizing treatment benefit as a weighted classification problem \citep{Zhang2012a,Rubin2012,Zhao2012,ZhouKosorork2015,ChenKosork2016}. 
 Several semi-parametric and non-parametric methods have been further proposed under this framework \citep{Zhao2012,Huang2014,Huang2015, Laber2015,Zhu2018}, and they are more robust against model misspecification than the parametric approaches. Under this framework, the convergence rates of the resulting estimators were thoroughly studied in \cite{Zhao2012} and \cite{zhang2018}, among others, but it is generally difficult to build statistical inference upon the estimated ITRs \citep{laberqian2019}. To solve this problem, \cite{jslz2019} proposed an entropy learning method, and \cite{wager2018} established asymptotic normality of a causal forest estimator.

The second general approach is to model the difference in average outcomes between two treatment groups conditional on predictive biomarkers that provide information about the effect of a therapeutic intervention. There are hybrids from the two approaches using mean-modeling \citep{Zhang2012a, taylor2015, zhang2018c, luckett}, but the hybrid methods also inherit the difficulty of developing inferential tools from the first general approach. In this paper, we focus on the idea of the second approach. With a single predictive biomarker, several authors have proposed to plot the estimated conditional treatment difference against the biomarker's values or its percentiles (e.g., a covariate-specific treatment
effect (CSTE) curve) obtained from nonparametric smoothing, along with its confidence bands, to derive  an optimal  ITR \citep{zhou2012bate, ma2014treatment, Janes2014, CSTE}. It provides a direct and effective visual way of using the predictive biomarker for deriving an ITR, but the nonparametric smoothing method suffers from ``curse of dimensionality". With multiple biomarkers,  \cite{Qian2011} considered a parametric conditional mean model that involves higher order terms, and employed a penalized method for estimation.  \cite{doi:10.1093/biostatistics/kxq060} proposed a two-step method, and \cite{Kang2014} devised a boosting iterative algorithm to reduce the bias caused by the model misspecification of a working model. Moreover, \cite{shi2018} proposed a penalized multi-stage A-learning, and \cite{zhu2019} developed a high-dimensional Q-learning method to simultaneously estimate the optimal dynamic treatment regimes and select important variables. 
The existing methods try to either provide a flexible modeling strategy by relaxing the restrictive structural assumption embodied in parametric models, or handle high-dimensional covariates, but not both. 

In our paper, we aim at developing a new method that estimates the outcome model in a flexible way as well as selecting important variables simultaneously, when a large number of baseline characteristics such as genetic variables are present. Moreover, we would like to provide a uniform inferential procedure that takes into account uncertainty associated with the estimated optimal ITR, and graphically explores heterogeneity of treatment effects based on varying levels of biomarkers. To achieve these goals, we propose a generalization of the CSTE curve suitable for high dimensional baseline covariates. The CSTE curve represents the predictive ability of covariates in evaluating whether a patient responds better to one treatment over another. It depicts the heterogeneous treatment effects on the outcome through an unknown function of covariates, and thus it can visually represent the magnitude of the predictive ability of the covariates.

We propose a penalized semi-parametric modeling approach for estimating the CSTE curve and selecting variables simultaneously. The proposed  semi-parametric model enjoys flexibility while achieving dimensionality reduction. It is motivated by the logistic varying-coefficient model considered in \citet{CSTE} for treatment selection with one covariate, and it meets the immediate needs from modern biomedical studies which can have a large number of baseline covariates. To make use of all covariates, one simple but effective way is to derive a weighted linear combination of all covariates as a summary predictor, and model the CSTE curve as an unknown function of this summary predictor. The weight or the coefficient for each covariate represents how important the covariate is for the prediction of the outcome. As a result, our proposed model involves two sets of high-dimensional coefficients fed into additive unknown functions for the two treatment groups. The development of the estimation procedure and the associated statistical properties is challenging and it needs different tools from the high-dimensional single-index model \citep{RADCHENKO2015266}. Moreover, we establish different convergence rates for the estimators of the high-dimensional coefficients and the estimators for the unknown additive functions, respectively. This new theoretical result makes it possible to further construct a simultaneous confidence band (SCB) for the CSTE curve based on the asymptotic extreme value distribution of a spline-backfitted kernel estimator \citep{MR2382655,doi:10.1080/01621459.2013.763726, Zheng2016} for the unknown function of interest, while \cite{RADCHENKO2015266} and other related works only provide a nonseparable convergence rate for the estimators of the coefficients and the unknown function in single-index models. Based on the SCB, we identify the subgroups of patients that benefit from each treatment, and the proposed method is flexible enough to depict both local and global associations between the treatment and baseline covariates.


The rest of the paper is organized as follows. In Section \ref{model-def},
we introduce the CSTE curve and the proposed semi-parametric logistic
single-index coefficient model. Section \ref{model-est} presents the
estimation procedure and the asymptotic properties of the proposed
estimators. Section \ref{selection} illustrates the application of the CSTE
curves and the SCBs for treatment selection. In Section \ref{simulation} we
evaluate the finite sample properties of the proposed method via simulation
studies, while Section \ref{real-world} illustrates the usefulness of the
proposed method through the analysis of a real data example. A discussion is
given in Section \ref{Discussion}. All technical proofs are relegated to the
online Supplemental Materials.

\section{Methodology}

\label{model-def} We consider a sample of $n$ subjects, a binary treatment,
denoted by $Z_{i}=1$ if the subject $i$ is assigned to treatment and $%
Z_{i}=0 $ otherwise, a $p$-dimensional vector of covariates, denoted by $%
X_{i}$, and binary-valued outcomes, denoted by $Y_{i}$. Let $\left(
Y_{i},Z_{i},X_{i}^{\top }\right) ,i=1,...,n$, be independent and identically
distributed (i.i.d.) copies of $\left( Y,Z,X\right) $. The goal is to  estimate the optimal treatment regime using the observed data. 
We consider the CSTE curve given in \cite{CSTE}, which has the following form:
\begin{equation*}
\mathrm{CSTE}(X)=\mathrm{logit}(E(Y(1)|X))-\mathrm{logit}(E(Y(0)|X)),
\end{equation*}%
where $\mathrm{logit}(u)=\log (u)-\log (1-u)$, and $Y(1)$ and $Y(0)$ denote the potential outcomes if the active treatment and the control treatment are received, respectively \citep{rubin2005}. Moreover, $Y\equiv ZY(1)+(1-Z)Y(0)$. Under the unconfoundeness
assumption such that $(Y(0),Y(1))\bot Z|X$, the CSTE curve can be re-expressed as
\begin{equation*}
\mathrm{CSTE}(X)=\mathrm{logit}(E(Y|X,Z=1))-\mathrm{logit}(E(Y|X,Z=0)).
\end{equation*} 

Denoting $\mu (X,Z)=E(Y|X,Z)$, we model the logarithm of odds ratio as
\begin{equation}
\mathrm{logit}(\mu (X,Z))=g_{1}(X^{\top }\beta _{1})\cdot Z+g_{2}(X^{\top
}\beta _{2}),  \label{semi-model}
\end{equation}
where $g_{1}(\cdot )$ and $g_{2}(\cdot )$ are unknown single-valued functions of $p$ variables, $\beta _{1}=(\beta _{11},\dots ,\beta _{1p})$ and $\beta _{2}=(\beta
_{21},\dots ,\beta _{2p})$ are two $p$-vectors of unknown parameters. We see 
\begin{equation*}
\mathrm{CSTE}(X)=\mathrm{logit}(\mu(X,1))-\mathrm{logit}(\mu(X,0))=g_{1}(X^{\top }\beta _{1}).
\end{equation*}
The two sets of high-dimensional coefficients and the two unknown functions in (\ref{semi-model}) are chosen to simultaneously maximize the log-likelihood function of the binomial distribution. The proposed model is a flexible semi-parametric model and is robust against model misspecification. We call it sparse logistic single index coefficient model (SLSICM). Our SLSICM contains the varying coefficient (VC) model considered in \cite{CSTE} as a special case. We have the same form when $p=1$. As an extension of the VC model, the varying-index coefficient model considered in \cite{VICM} can be applied to the cases with several covariates and continuous responses. Moreover, an important and related work, \cite{song2017}, proposed a semi-parametric model in which $g_1$ has a single-index structure and $g_2$ is a pure nonparametric function of $X$. This model suffers from the curse of dimensionality when the number of covariates is large. 
In our paper, we allow the number of covariates to be much larger than sample size. As a result, our proposed SLSICM involves two sets of high-dimensional unknown coefficients built into additive unknown functions with a nonlinear link function, and it is considered as a hybrid of the high-dimensional single index model (HSIM) and the nonparametric additive model. Developing the estimation procedure and the statistical theories for SLSICM is challenging, and it needs different tools from HSIM. It is worth noting that estimation of HSIM has been studied in the past several years, and most existing works use a sliced inverse regression approach to estimate the index coefficients in HSIM under a linearity condition on the covariates; see \cite{JL2014} and \cite{NLC2016}, among others. This method is not applicable to our setting, and it does not directly estimate the unknown function. \cite{RADCHENKO2015266} proposed to estimate the coefficients and the unknown function in HSIM jointly, but they only provide a nonseparable convergence rate for the resulting estimators of the coefficients and the unknown function. In our proposed SLSICM, we establish different convergence rates for the estimators of the coefficients $\beta_1$ and $\beta_2$ and the estimators of the unknown functions $g_1$ and $g_2$, respectively. This new theoretical result makes it possible to further construct the asymptotic SCB based on the asymptotic extreme value distribution of a spline-backfitted kernel estimator for the CSTE curve.

Unlike parametric models, the parameter vectors $\beta_k$ are not identified
without further assumptions. For the purpose of model identification, we
assume that $\beta_k$ for $k=1,2$ belong to the parameter space $\Theta =\{
\beta=(\beta_1^{\top },\beta_{2}^{\top })^{\top }:\Vert\beta_{k}\Vert
=1,\beta _{k1}>0,\beta_{k}\in R^{p}, k=1,2 \}, $ where $\Vert \cdot \Vert $
denotes the $l_{2}$ norm of a vector. This restriction together with Assumption 2 given in Section \ref{theory} will make the model (\ref{semi-model}) identifiable. We will provide a formal proof of the model identifiability in Section 1.1 of the Supplemental Materials.  Based on the constraint that $\Vert \beta _{k}\Vert =1$, we
eliminate the first component in $\beta _{k}$ and obtain the resulting
parameter space: $\Theta _{-1}=\{\beta _{k,-1}=(\beta _{k2},\dots ,\beta
_{kp})^{\top }:\sum\nolimits_{j>1}\beta _{kj}^{2}<1,k=1,2\}$. Let $\beta
_{k1}=\sqrt{1-\sum\nolimits_{j>1}\beta _{kj}^{2}}$. The derivative with respect
to the coefficients $(\beta_{k2},\dots ,\beta_{kp})^{\top}$ can be easily
obtained using the chain rule under the above parameter space. For
high-dimensional problems (with a large number of covariates), $p$ can be
much larger than $n$ but only a small number of covariates are important or
relevant for treatment selection. To this end, we assume that the number of
nonzero elements increases as $n$ increases, but it is much smaller than $n$%
. Without loss of generality, we assume that only the first $s_k=s_{kn}$
components of $\beta_k$ are non-zeros, i.e., we can write the true values as
$\beta_{k}$ $=(\beta_{k1},\dots, \beta_{ks_k}, 0, \dots,0)^{\top }$. For model identifiability, we require one component in $\beta_k$ to be nonzero. Without loss of generality, we let the first component $\beta_{k1}$ to be nonzero.

\section{Estimation and Theory}

\label{model-est}

\subsection{Algorithm}

\label{sparse-logit} In this subsection, we discuss the estimation of
model (\ref{semi-model}). We minimize the negative log-likelihood function
simultaneously with respect to the parameters $\beta _{k}$ and the functions
$g_{k}$ for $(k=1,2)$. The SLSICM has the form $\mathrm{logit}(\mu
(X,Z))=g_{1}(X^{\top }\beta _{1})\cdot Z+g_{2}(X^{\top }\beta _{2}).$
Therefore, we seek the minimizer of the following negative log-likelihood function
given $(X_{i},Y_{i},Z_{i}),i=1,\dots ,n$:
\begin{equation}
\frac{1}{n}\sum_{i=1}^{n}\log \{1+\exp (g_{1}(X_{i}^{\top }\beta
_{1})Z_{i}+g_{2}(X_{i}^{\top }\beta _{2}))\}-\frac{1}{n}\sum_{i=1}^{n}Y_{i}%
\{g_{1}(X_{i}^{\top }\beta _{1})Z_{i}+g_{2}(X_{i}^{\top }\beta _{2})\}.
\label{EQ:Ln}
\end{equation}%
To overcome the problem of high-dimensional covariates, we exploit the
sparsity through parameter regularization. With the sparsity constraint of $%
\beta _{k}$'s, we minimize the following penalized negative log-likelihood
\begin{equation}
\frac{1}{n}\sum_{i=1}^{n}\log \{1+\exp (g_{1}(X_{i}^{\top }\beta
_{1})Z_{i}+g_{2}(X_{i}^{\top }\beta _{2}))\}-\frac{1}{n}\sum_{i=1}^{n}Y_{i}%
\{g_{1}(X_{i}^{\top }\beta _{1})Z_{i}+g_{2}(X_{i}^{\top }\beta
_{2})\}+\sum_{k=1}^{2}\sum_{j=2}^{p}p(\beta _{kj},\lambda ),
\label{EQ:Ln:sparse}
\end{equation}%
where $p(\cdot )$ is a penalty function with a tuning parameter $\lambda $
that controls the level of sparsity in $\beta _{k}$, $k=1,2$.

The functions $g_{k}(\cdot )$, $k=1,2$, are unspecified and are estimated
using B-splines regression. Next, we introduce the B-splines that will be
used to approximate the unknown functions. For $k=1,2$, we assume the support
of $g_{k}(\cdot )$ is $[\inf_{X}(X^{\top }\beta _{k}),\sup_{X}(X^{\top
}\beta _{k})]=[a_{k},b_{k}].$ Let $a_{k}=t_{0,0}<t_{1,k}<\cdots
<t_{N_{k},k}<b_{k}=t_{N_{k}+1}$ be an equally-spaced partition of $%
[a_{k},b_{k}]$, called interior knots. Then, $[a_{k},b_{k}]$ is divided into
subintervals $I_{\ell ,k}=[t_{\ell ,k},t_{\ell +1,k})$, $0\leq \ell \leq
N_{k}-1$ and $I_{N_{k}}=[t_{N_{k}},t_{N_{k}+1}]$, satisfying $\max_{0\leq
\ell \leq N_{k}}|t_{\ell +1,k}-t_{\ell ,l}|/\min_{0\leq \ell \leq
N_{k}}|t_{\ell +1,k}-t_{\ell ,k}|\leq M$ uniformly in $n$ for some constant $%
0<M<\infty $, where $N_{k}\equiv N_{k,n}$ increases with the sample size $n$%
. We write the normalized B spline basis of this space \citep{MR1900298} as $%
B_{k}(u_{k})=\{B_{\ell ,k}(u_{k}):1\leq \ell \leq L_{n,k}\}^{\top }$, where
the number of spline basis functions is $L_{n,k}=L_{k}=N_{k}+q_{k}$, and $q_{k}$
is the spline order. For computational convenience, we let $N_{k}=N$ and $%
q_{k}=q$ so that $L_{k}=L$. In practice, cubic splines with order $q=4$ are
often used. By the result in \citet{MR1900298}, the nonparametric function
can be approximated well by a spline function such that $g_{k}(X^{\top
}\beta _{k})\approx B_{k}(X^{\top }\beta _{k})^{\top }\delta _{k}$ for some $%
\delta _{k}\in R^{L}$, $k=1,2.$ For notational simplicity, we write $%
B_{k}(X^{\top }\beta _{k})$ as $B(X^{\top }\beta _{k})$. Therefore, the
estimates of the unknown index parameters $\beta _{k}$ and the spline coefficients $\delta _{k}$ are the minimizers of
\begin{eqnarray}
&&\frac{1}{n}\sum_{i=1}^{n}\log \{1+\exp (B_{1}(X_{i}^{\top }\beta
_{1})^{\top }\delta _{1}Z_{i}+B_{2}(X_{i}^{\top }\beta _{2})^{\top }\delta
_{2})\} \notag\\
&&-\frac{1}{n}\sum_{i=1}^{n}Y_{i}\{B_{1}(X_{i}^{\top }\beta _{1})^{\top
}\delta _{1}z_{i}+B_{2}(X_{i}^{\top }\beta _{2})^{\top }\delta
_{2}\}+\sum_{k=1}^{2}\sum_{j=2}^{p}p(\beta _{kj},\lambda ),
\label{EQ:Ln:sparse2}
\end{eqnarray}%

Denote $U_{k}=X^{\top }\beta _{k}$ for $k=1,2$. For notational simplicity,
we use $U$ to denote $U_{k}$ by suppressing $k$. To allow for possibly
unbounded support $\mathcal{U}$ of $U$, according to the method given in
\cite{CC15}, we can weight the B-spline basis functions by a sequence of
nonnegative weighting functions $w_{n}:\mathcal{U}\rightarrow \{0,1\}$ given
by $w_{n}(u)=1$ if $u\in \mathcal{D}_{n}$ and $w_{n}(u)=0$ otherwise, where $%
\mathcal{D}_{n}\subseteq \mathcal{U}$ is compact, convex and has nonempty
interior and $\mathcal{D}_{n}\subseteq \mathcal{D}_{n+1}$ for all $n$. For
the choices of $\mathcal{D}_{n}$, we refer to \cite{CC15} for the
discussions. When $\mathcal{U}$ is compact, we simply set $%
w_{n}(u)=1$ for all $u\in \mathcal{U}$. Then we obtain a set of new B-spline
basis functions: {\normalsize $B_{k}^{w}(u_{k})=\{B_{\ell
,k}(u_{k})w_{n}(u_{k}):1\leq \ell \leq L_{n,k}\}^{\top }$, }and replace
$B_{k}(u_{k})$ by $B_{k}^{w}(u_{k})$ in the above minimization
problem for obtaining the estimators  $\widehat{\beta }_{k}$ and $\widehat{%
\delta }_{k}$. As given in \cite{CC15}, the asymptotic properties including
consistency and asymptotic distribution for the estimator of the
unknown function $g_1(u)$ still hold for $u\in \mathcal{D}_{n}$.

In principle, we would like to obtain the estimates of $\beta _{k}$ and $\delta _{k}$  by minimizing the penalized negative log-likelihood given in (\ref{EQ:Ln:sparse2}). In practice, this minimization is different to achieve, and thus an iterative algorithm \citep{WE83} is often applied. To this end, we obtain the estimates of  $\widehat{\beta }_{k}$ and $\widehat{\delta }_{k}$  through an iterative algorithm described as follows, although our theoretical properties given in Section \ref{theory} are established for the estimators which minimize (\ref{EQ:Ln:sparse2}).

\noindent \textbf{Step 1}: Given $\beta _{k}$, the solution of $\delta _{k}$
is easily obtained. Reexpressing the model gives
\begin{equation*}
\mathrm{logit}(\mu (X,Z))\approx B(X^{\top }\beta _{k})^{\top }\delta
_{1}Z+B(X^{\top }\beta _{k})^{\top }\delta _{2}=\left( ZB(X^{\top }\beta
_{k})^{\top },B(X^{\top }\beta _{k})^{\top }\right) \left(
\begin{array}{c}
\delta _{1} \\
\delta _{2}%
\end{array}%
\right) .
\end{equation*}%
This can be viewed as a logistic regression model using $\left( ZB(X^{\top }\beta
_{k})^{\top },B(X^{\top }\beta _{k})^{\top }\right) ^{\top }$ as the
regressors without intercept term.

\noindent \textbf{Step 2}: Given $\delta _{k}$, it remains to find the
solution that minimizes (\ref{EQ:Ln:sparse}) with respect to $\beta _{k}$.
Let $\beta _{k}^{old}$ and $\beta _{k,-1}^{old}$ be the current estimates
for $\beta _{k}$ and $\beta _{k,-1}$, respectively. Let $\widetilde{g}%
_{k}(X^{\top }\beta _{k})=B_{k}(X^{\top }\beta _{k})^{\top }\delta _{k}$. We
approximate
\begin{equation*}
\widetilde{g}_{k}(X^{\top }\beta _{k})\approx \widetilde{g}_{k}(X^{\top
}\beta _{k}^{old})+\widetilde{g}_{k}^{\prime }(X^{\top }\beta
_{k}^{old})X^{\top }J(\beta _{k}^{old})(\beta _{k,-1}-\beta _{k,-1}^{old}),
\end{equation*}%
where $J(\beta _{k})=\partial \beta _{k}/\partial \beta _{k,-1}=(-\beta
_{k,-1}/\sqrt{1-\rVert \beta _{k,-1}\lVert _{2}^{2}},I_{p-1})^{\top }$ is
the Jacobian matrix of size $p$ by $p-1$. To obtain the sparse estimates of $%
\beta _{k}$, we carry out a regularized logistic regression with $\left[ \{Z%
\widetilde{g}_{1}^{\prime }(X^{\top }\beta _{1}^{old})J(\beta
_{1}^{old})^{\top }X\}^{\top },\{\widetilde{g}_{2}^{\prime }(X^{\top }\beta
_{2}^{old})J(\beta _{2}^{old})^{\top }X\}^{\top }\right] ^{\top }$ as the
regressors with a known intercept term given as
\begin{equation*}
Z\widetilde{g}_{1}(X^{\top }\beta _{1}^{old})+\widetilde{g}_{2}(X^{\top
}\beta _{2}^{old})-Z\widetilde{g}_{1}^{\prime }(X^{\top }\beta
_{1}^{old})X^{\top }J(\beta _{1}^{old})\beta _{1,-1}^{old}-\widetilde{g}%
_{2}^{\prime }(X^{\top }\beta _{2}^{old})X^{\top }J(\beta _{2}^{old})\beta
_{2,-1}^{old}.
\end{equation*}%
This produces an updated vector $\beta _{k,-1}^{new}$. Then we set $\beta
_{k}^{new}=(\sqrt{1-||\beta _{k,-1}^{new}||^{2}},(\beta _{k,-1}^{new})^{\top
})^{\top }$, for $k=1,2$. Steps 1 and 2 are repeated until convergence.

We obtain the initial value of $\beta _{k}$ through fitting a regularized
logistic regression by assuming that $g_{k}(X^{\top }\beta _{k})=X^{\top
}\beta _{k}$. In step 2, we use the coordinate descent algorithm %
\citep{MR2810396} to fit the regularized regression. Moreover, we choose to
use the non-convex penalties such as MCP and SCAD which induce nearly
unbiased estimators. The MCP \citep{MR2604701} has the form $p_{\gamma
}(t,\lambda )=\lambda \int_{0}^{t}(1-x/(\gamma \lambda ))_{+}dx,\gamma >1$
and the SCAD \citep{MR1946581} penalty is $p_{\gamma }(t,\lambda )=\lambda
\int_{0}^{t}\min \{1,(\gamma -x/\lambda )_{+}/(\gamma -1)\}dx,\gamma >2,$
where $\gamma $ is a parameter that controls the concavity of the penalty
functions. In particular, both penalties converge to the $L_{1}$ penalty as $%
\gamma \rightarrow \infty $. We put $\gamma $ in the subscript to indicate
the dependence of these penalty functions on it. In practice, we treat $%
\gamma $ as a fixed constant. The B-spline basis functions and their
derivatives are calculated using the \verb|bsplineS| function in R package
\verb|fda|.

From the above algorithm, we obtain the spline estimators of the functions $%
g_{1}(\cdot )$ and $g_{2}(\cdot )$. However, the spline estimator only has
convergence rates but its asymptotic distribution is not available in the additive model settings with multiple unknown functions \citep{S85}, so no measures
of confidence can be assigned to the estimators for conducting statistical
inference\citep{MR2382655}. The spline-backfitted kernel (SBK) estimator is designed to overcome this issue for generalized additive models %
\citep{doi:10.1080/01621459.2013.763726}, which combines the strengths of
kernel and spline smoothing, is easy to implement and has asymptotic
distributions. Denote the SBK estimator of $g_{1}(\cdot )$ as $\widehat{g}_{1,sbk}(\cdot)$. We obtain $\widehat{g}_{1,sbk}(X_{0}^{\top }\widehat{\beta }%
_{1})$ for a new input vector $X_{0}$ as follow:

\noindent \textbf{Step 3}: Given the spline estimate $\widehat{g}%
_{2}(X_{i}^{\top }\widehat{\beta }_{2})$, the loss criterion for a local
linear logistic regression can be expressed as the following negative
quasi-likelihood function:
\begin{equation*}
l(a,b,X) =-\frac{1}{n}\sum_{i}[Y_{i}(aZ_{i}+\widehat{g}%
_{2}(X_{i}^{\top }\widehat{\beta }_{2})) -\log (1+\exp (aZ_{i}+\widehat{g}_{2}(X_{i}^{\top }\widehat{\beta}%
_{2})))]K_{h}(X_{i}^{\top }\widehat{\beta }_{1}-X^{\top }\widehat{\beta }%
_{1}),
\end{equation*}
where $K_{h}(\cdot )$ is a kernel function with bandwidth $h$. We obtain the
estimate $\widehat{a}(X_{0})$ by minimizing the above loss function. Then the
predicted CSTE value at $X_{0}$ is
\begin{equation}
\mathrm{CSTE}(X_{0})\approx \widehat{g}_{1,sbk}(X_{0}^{\top }\widehat{\beta }%
_{1})=\widehat{a}(X_{0}).  \label{SBK1}
\end{equation}%
Based on the above SBK\ estimator of CSTE, we can construct a SCB which is used for optimal treatment selection. The details
will be discussed in Section \ref{selection}.

\subsection{Asymptotic Analysis}

\label{theory} For any positive sequences
$\{a_{n}\}$ and $\{b_{n}\}$, let $a_{n}\asymp b_{n}$ denote $%
\lim_{n\rightarrow \infty }a_{n}b_{n}^{-1}=C$ for a constant $0<C<\infty $
and $a_{n}\ll b_{n}$ denote $\lim_{n\rightarrow \infty }a_{n}b_{n}^{-1}=0$.
For a vector $a=(a_{1},\dots ,a_{p})^{\top }\in R^{p},$ denote $\Vert a\Vert
=(\sum_{l=1}^{p}a_{l}^{2})^{1/2}$ and $\Vert a\Vert _{\infty
}=\max_{l}|a_{l}|$. For a matrix $A=(A_{ij})$, denote $\Vert A\Vert
=\max_{\Vert \zeta \Vert =1}\Vert A\zeta \Vert $, $\Vert A\Vert _{\infty
}=\max_{i}\sum_{j}|A_{ij}|$, and $\Vert A\Vert _{2,\infty
}=\max_{i}||A_{i}|| $, where $A_{i}$ is the $i$th row. For a symmetric
matrix $A$, let $\lambda _{\min }(A)$ and $\lambda _{\max }(A)$ be the
smallest and largest eigenvalues of $A$, respectively. We assume that the
nonsparsity size $s=\max (s_{1},s_{2})\ll n$ and the dimensionality
satisfies $\log p=O(n^{\alpha })$ for some $\alpha \in (0,1)$. Denote $\beta
_{k1}=(\beta _{k1},\dots ,\beta _{ks_{k}})^{\top },\beta _{k1,-1}=(\beta
_{k2},\dots ,\beta _{ks_{k}})^{\top },\beta _{k2}=(\beta _{k(s_{k}+1)},\dots
,\beta _{kp})^{\top }.$ Then we write $\beta _{(1)}=(\beta _{11}^{\top
},\beta _{21}^{\top })^{\top },\beta _{(2)}=(\beta _{12}^{\top },\beta
_{22}^{\top })^{\top },\beta _{(1),-1}=(\beta _{11,-1}^{\top },\beta
_{21,-1}^{\top })^{\top }.$ Denote the Jacobian matrix as $J(\beta
_{k1})=\partial \beta _{k1}/\beta _{k1,-1},k=1,2$, and $J(\beta
_{(1)})=\partial \beta _{(1)}/\partial \beta _{(1),-1}=diag(J(\beta
_{11}),J(\beta _{21})),$ which is a block diagonal matrix. We use the
superscript `0' to represent the true values.

Denote the first $s_{k}(1\leq k\leq 2)$ components of $X_{i}$ as $%
X_{i,k1}=(x_{ij},1\leq j\leq s_{k})^{\top }$ and the last $p-s_{k}$
components as $X_{i,k2}=(x_{ij},s_{k}<j\leq p)^{\top }.$ Denote $%
S(x)=(1+e^{-x})^{-1}$ as the sigmoid function. Then the true expected value
of the response given $(X,Z)$ is
\begin{equation*}
E(Y|X,Z)=\mu (X,Z)=S(g_{1}(X^{\top }\beta _{1})Z+g_{2}(X^{\top }\beta _{2})).
\end{equation*}%
Define the space $\mathcal{M}$ as a collection of functions with finite L$%
_{2}$ norm on $\mathscr{C}\times \{0,1\}$ by%
\begin{equation*}
\mathcal{M=}\left\{ g(x,z)=g_{1}(x^{\top }\beta _{1}^{0})z+g_{2}(x^{\top
}\beta _{2}^{0}),E\{g_{k}(X^{\top }\beta _{k}^{0})\}^{2}<\infty \right\} .
\end{equation*}%
For a given random variable $U$, define its projection onto the space $%
\mathcal{M}$ as
\begin{equation*}
\mathbb{P}_{\mathcal{M}}(U)=\arg \min_{g\in \mathcal{M}}E\left[
w^{0}\{U-g(X,Z)\}^{2}\right] ,
\end{equation*}%
where $w^{0}=\pi ^{0}(1-\pi ^{0})$ and $\pi ^{0}=\mu (X,Z)$. For a vector $%
U=\{U_{1},...,U_{d}\}$, let
\begin{equation*}
\mathbb{P}_{\mathcal{M}}(U)=\{\mathbb{P}_{\mathcal{M}}(U_{1}),...,\mathbb{P}%
_{\mathcal{M}}(U_{d})\}^{\top }.
\end{equation*}
Moreover, we define
\begin{equation}
\Omega _{n}=\frac{1}{n}\sum_{i=1}^{n}J(\beta _{(1)}^{0})^{\top }\left(
\begin{array}{c}
g_{1}^{\prime }(X_{i,11}^{\top }\beta _{11}^{0})\widetilde{X}_{i,11}Z_{i} \\
g_{2}^{\prime }(X_{i,21}^{\top }\beta _{21}^{0})\widetilde{X}_{i,12}%
\end{array}%
\right) \left(
\begin{array}{c}
g_{1}^{\prime }(X_{i,11}^{\top }\beta _{11}^{0})\widetilde{X}_{i,11}Z_{i} \\
g_{2}^{\prime }(X_{i,21}^{\top }\beta _{21}^{0})\widetilde{X}_{i,12}%
\end{array}%
\right) ^{\top }J(\beta _{(1)}^{0}),  \label{EQ:phin}
\end{equation}%
where $\widetilde{X}_{i,kv}=X_{i,kv}-\mathbb{P}_{\mathcal{M}}(X_{i,kv})$ for
$k,v=1,2$, and
\begin{equation*}
\Phi _{n}=\frac{1}{n}\sum_{i=1}^{n}\sigma ^{2}(X_{i},Z_{i})J(\beta
_{(1)}^{0})^{\top }\left(
\begin{array}{c}
g_{1}^{\prime }(X_{i,11}^{\top }\beta _{11}^{0})\widetilde{X}_{i,11}Z_{i} \\
g_{2}^{\prime }(X_{i,21}^{\top }\beta _{21}^{0})\widetilde{X}_{i,21}%
\end{array}%
\right) \left(
\begin{array}{c}
g_{1}^{\prime }(X_{i,11}^{\top }\beta _{11}^{0})\widetilde{X}_{i,11}Z_{i} \\
g_{2}^{\prime }(X_{i,21}^{\top }\beta _{21}^{0})\widetilde{X}_{i,21}%
\end{array}%
\right) ^{\top }J(\beta _{(1)}^{0}),
\end{equation*}%
where $\sigma ^{2}(X,Z)=E[\{Y-S(g(X,Z))\}^{2}|X,Z]$ .

To establish asymptotic properties, we need the following regularity
conditions.

\begin{assumption}
The penalty function $p_{\gamma}(t,\lambda)$ is a non-decreasing symmetric
function and concave on $[0,\infty)$. For some constant $a>0$, $%
\rho(t)=\lambda^{-1}p_{\gamma}(t,\lambda)$ is a constant for all $t\geq
a\lambda$ and $\rho(0)=0$. $\rho^{\prime}(t)$ exists and is continuous
except for a finite number of $t$ and $\rho^{\prime}(0+)=1$. \label{A1}
\end{assumption}
\begin{assumption}
For $k=1,2$, for any given $\beta _{k}\in $ $\Theta $, $g_{k}$ is a nonconstant function on $\{x\in {\mathcal X\}}$, where $\mathcal{X}$ is compact, convex and has nonempty
interior, and $g_{k}\in \mathcal{H}_{r}$ for some $r>1$%
, where $\mathcal{H}_{r}$ is the collection of all functions such
that the $q^{\text{th}}$ order derivative satisfies the H\"{o}lder condition
of order $\gamma $ with $r\equiv q+\gamma $ and $q\geq 1$, i.e. for any $\phi \in \mathcal{%
H}_{r}$, there is a $C_{0}\in (0,\infty )$ such that $|\phi
^{(q)}(u_{1})-\phi ^{(q)}(u_{2})|\leq C_{0}|u_{1}-u_{2}|^{\gamma }$
for all $u_{1}$ and $u_{2}$ in the support of $g_{k}$. \label{A2}
\end{assumption}

\begin{assumption}
For $1\leq j\leq s_{k}$, $E(X_{j}^{2+2(\varkappa +1)})\leq C_{\varkappa ,k}$
for some constant $\varkappa \in (0,1)$. For $s_{k}\leq j\leq p-s_{k}$, $%
E(|X_{j}|^{2+\varrho })\leq C_{\varrho ,k}$, for some $\varrho
>(8/3)(1-\alpha )^{-1}-2$ and $\varrho \geq 2$. \label{A3}
\end{assumption}

\begin{assumption}
There exist $c,c^{\prime },c_{k}\in (0,\infty )$ such that $\lambda _{\min
}(\Omega _{n})\geq c$, $\lambda _{\min }(\Phi _{n})\geq c^{\prime }$ almost
surely, and $\Vert E(\widetilde{X}_{k2}\widetilde{X}_{k1}^{\top })\Vert
_{2,\infty }\leq c_{k}$, where $\widetilde{X}_{kv}=X_{kv}-\mathbb{P}_{%
\mathcal{M}}(X_{kv})$, for $k,v=1,2$. \label{A4}
\end{assumption}

\begin{assumption}
Let $w_{n}=2^{-1}\min \{|\beta _{kj}^{0}|:2\leq j\leq s,k=1,2\}.$ Assume
that $\max ((s/n)^{1/2},\newline  L^{1-r}+L^{3/2}n^{\alpha /2-1/2})\ll \lambda \ll
w_{n}$. \label{A5}
\end{assumption}

Assumption \ref{A1} is a typical condition on the penalty function, see \cite%
{FLv11}. The concave penalties such as SCAD and MCP satisfy Assumption \ref%
{A1}. Assumption \ref{A2} is a typical smoothness condition on the unknown
nonparametric function, see for instance Condition (C3) in \cite%
{6cfe1e56953040a0bb6a2f0f4c1310c4}. Moreover, we require that the unknown functions be nonconstant functions in order to identify the index parameters $\beta _{k}$. Assumption \ref{A3} is required for the
covariates, see Condition (A5) in \cite{MR2780816}. Moreover, the design
matrix needs to satisfy Assumption \ref{A4}. A similar condition can be
found in \cite{FLv11}. Assumption \ref{A5} assumes that half of the minimum
nonzero signal in $\beta_k^{0}$ is bounded by some thresholding value, which
is allowed to go to zero as $n\rightarrow\infty$. This assumption is needed
for variable selection consistency established in Theorem \ref{th1}.

Denote $\beta _{-1}=(\beta _{11,-1}^{\top },\beta _{12}^{\top },\beta
_{21,-1}^{\top },\beta _{22}^{\top })^{\top }$. Let $s=\max (s_{1},s_{2})$.
Theorem \ref{th1} establishes the consistency for the parameters in model (%
\ref{semi-model}).

\begin{theorem}
Under Assumptions (A1)-(A5), and $\alpha \in (0,(2r-5/2)/(2r-1))$, $%
L^{1-r}s^{1/2}=o(1)$, $n^{\alpha -1}(L^{3}+s)=o(1)$, $n^{1/2}L^{1-2r}=o(1)$,
$\log (n)(s^{1/2}+L^{1/2})L^{3/2}n^{-1/2}=o(1)$, there exists a strict local
minimizer $\widehat{\beta }_{-1}=(\widehat{\beta }_{11,-1}^{\top },\widehat{%
\beta }_{12}^{\top },\widehat{\beta }_{21,-1}^{\top },\widehat{\beta }%
_{22}^{\top })^{\top }$ of the loss function given in (\ref{EQ:Ln:sparse})
such that $\widehat{\beta }_{k2}=0$ for $1\leq k\leq 2$ with probability
approaching $1$ as $n\rightarrow \infty $, and $\Vert \widehat{\beta }%
_{-1}-\beta _{-1}^{0}\Vert =O_{p}(\sqrt{s/n})$. \label{th1}
\end{theorem}

\textbf{Remark. }Based on the assumption given in Theorem \ref{th1}, the
number of spline basis functions needs to satisfy $n^{1/\{2(2r-1)\}}\ll L\ll
\min \{n^{1/4}\{\log (n)\}^{-1},n^{(1-\alpha )/3}\}$.

The following Theorem presents the convergence rate for the spline estimator
of the unknown functions.

\begin{theorem}
Under conditions given in Theorem \ref{th1}, we have $n^{-1}\sum_{i=1}^{n}\{%
\widehat{g}_k(X_{i}^{\top}\widehat{\beta}_k)-{g}_k(X_{i}^{\top}\beta_k^{0})%
\}^{2} = O_{p}(L^{-2r}+L/n+s/n)$, for $k=1,2$. \label{conv}
\end{theorem}

To estimate the SCB, we need the following
assumptions, see Assumptions (A4) and (A6) in \cite{Zheng2016}.

\begin{assumption}
Let $r=2$. The kernel function $K$ is symmetric probability density function
supported on $[-1,1]$ and has bounded derivative. The bandwidth $h$
satisfies $h=h_{n}=o(n^{-1/5}(\log n)^{-1/5})$ and $h^{-1}=O(n^{1/5}(\log
n)^{\delta })$ for some constant $\delta >1/5$. \label{A6}
\end{assumption}

\begin{assumption}
The joint density of $X^{\top}\beta_1^0$ and $X^{\top}\beta_2^0$ is a bounded and continuous function. The marginal probability density functions have continuous derivatives and the same bound as the joint density. When $\beta_1^0=\beta_2^0=\beta^0$, the probability density function of $X^{\top}\beta^0$ is bounded, continuous, and has continuous derivative.  
\label{A7}
\end{assumption}

We borrow some notations in \cite{Zheng2016}:
\begin{equation}
\sigma ^{2}(u)=E[\mu (u,Z)(1-\mu (u,Z))|Z=1],D(u)=f(u)\sigma
^{2}(u),v^{2}(u)=||K||_{2}^{2}f(u)\sigma ^{2}(u),  \label{sigma-v-d}
\end{equation}%
where $\mu (u,Z)=S(g_{1}(u)Z+g_{2}(X^{\top }\beta _{2}^{0}))$ and $f(u)$ is
the density function of $X^{\top }\beta _{1}^{0}$. Define the quantile
function
\begin{equation*}
Q_{h}(\alpha )=a_{h}+a_{h}^{-1}[\log (\sqrt{C_{K}}/(2\pi ))-\log (-\log
\sqrt{1-\alpha })]
\end{equation*}%
for any $\alpha \in (0,1)$, where $a_{h}=\sqrt{-2\log h}$ and $%
C_{K}=||K^{\prime }||_{2}^{2}/||K||_{2}^{2}$.  Without loss of generality,
assume $x^{\top }\beta _{1}^{0}$ is in the range $[0,1]$ and let $\mathscr{C}
$ be a set of $x$ such that $\mathscr{C}=\{x:~x^{\top }\beta _{1}^{0}\in
\lbrack h,1-h],x\in R^{p}\}$. Theorem \ref{band:thm} is an adaptation from
Theorem 1 in \cite{Zheng2016}. It provides a method to construct the
SCB for $g_{1}$. 

\begin{theorem}
\label{band:thm} Under Assumptions \ref{A1}-\ref{A7}, and $\alpha \in (0,2/5)
$, $sn^{-1/10}(\log n)^{-3/5}=O(1)$ and $n^{1/5}\ll L\ll \min
\{n^{1/4}\{\log (n)\}^{-1},n^{(1-\alpha )/3}\}$, we have
\begin{equation*}
\lim_{n\rightarrow \infty }P\left( \sup_{x\in \mathscr{C}}\left\vert \frac{%
\widehat{g}_{1,sbk}(x^{\top }\widehat{\beta }_{1})-g_{1}(x^{\top }\beta
_{1}^{0})}{\sigma _{n}(x^{\top }\widehat{\beta }_{1})}\right\vert \leq
Q_{h}(\alpha )\right) =1-\alpha ,
\end{equation*}%
where $\sigma _{n}(x^{\top }\beta _{1})=n^{-0.5}h^{-0.5}v(x^{\top }\beta
_{1})/D(x^{\top }\beta _{1})$. 
\end{theorem}

\textbf{Remark 1. }Based on the proof for Theorem \ref{band:thm}, we can readily obtain that \newline $\sigma _{n}(x^{\top }\widehat{\beta }%
_{1})^{-1}\{\widehat{g}_{1,sbk}(x^{\top }\widehat{\beta }_{1})-g_{1}(x^{\top
}\beta _{1}^{0})\}\rightarrow \mathcal{N}(0,1)$ in distribution, for given $%
x\in \mathscr{C}$. Thus, the {\normalsize $100(1-\alpha )\%$ pointwise
confidence interval for $g_{1}(x^{\top }\beta _{1}^{0})$ is $\widehat{g}%
_{1,sbk}(x^{\top }\widehat{\beta }_{1})\pm Z_{1-\alpha /2}\sigma
_{n}(x^{\top }\widehat{\beta }_{1})$, where }$Z_{1-\alpha /2}$ is the $%
(1-\alpha /2)100\%$ quantile of the standard normal distribution.

\textbf{Remark 2. }For details of implementations of kernel and spline
estimation for functions in (\ref{sigma-v-d}), we refer to \cite{Zheng2016}.
As suggested in \cite{Zheng2016}, we use a data-driven under smoothing
bandwidth $h=h_{opt}(\log n)^{-1/4}$ and $h_{opt}$ is given in \cite%
{Zheng2016}. We let the number of spline interior knots be $\left\lfloor
n^{1/5}(\log n)\right\rfloor +1$. The $100(1-\alpha )\%$ SCB for $g_{1}(x^{\top }\beta _{1}^{0})$ is $\widehat{g}%
_{1,sbk}(x^{\top }\widehat{\beta }_{1})\pm \sigma _{n}(x^{\top }\widehat{%
\beta }_{1})Q_{h}(\alpha ).$

\textbf{Remark 3. } From Theorem \ref{band:thm}, we see that the width of the SCB has the order of $\sqrt{\log (h)n/h}$. The width of the pointwise confidence interval given in Remark 1 has the order of $\sqrt{n/h}$. Based on the assumption for the bandwidth $h$, the SCB is wider than the pointwsie confidence interval asymptotically. In deriving the asymptotic distribution of the SBK estimator for the unknown function, it also involves the estimation error from the parameter estimation, which is given in Theorem \ref{th1}. We assume that $s$ increases with $n$ in a slow rate, so that this estimation error is negligible in the construction of the asymptotic SCB. Our interest focuses on constructing the SCB for the CSTE curve, based on which we make treatment recommendations to a group of new patients. For conducting inference for the parameters, it can be a future interesting research topic to explore. 

The theoretical result in Theorem \ref{band:thm} is obtained from the asymptotic extreme value distribution of the SBK estimator for the CSTE curve. This is the key result for obtaining our asymptotic SCB, which achieves the nominal confidence level asymptotically. It is worth noting that we provide different convergence rates for the estimators of the high-dimensional coefficients and the estimators for the unknown functions in Theorems \ref{th1} and \ref{conv}, respectively. This theoretical result enables us to further derive the asymptotic distribution of $\widehat{g}_{1,sbk}$,  when the number of important covariates satisfies certain order given in Theorem \ref{band:thm}. 
The SCB provides an important uniform inferential tool for identifying subgroups of patients that benefit from each treatment and make treatment recommendations to the subgroups, while the pointwise confidence interval can only provide treatment selection for a given patient.

\section{Treatment Selection via Confidence Bands}

\label{selection}

\begin{figure}[hbt]
\centering
\includegraphics[width=0.8\textwidth]{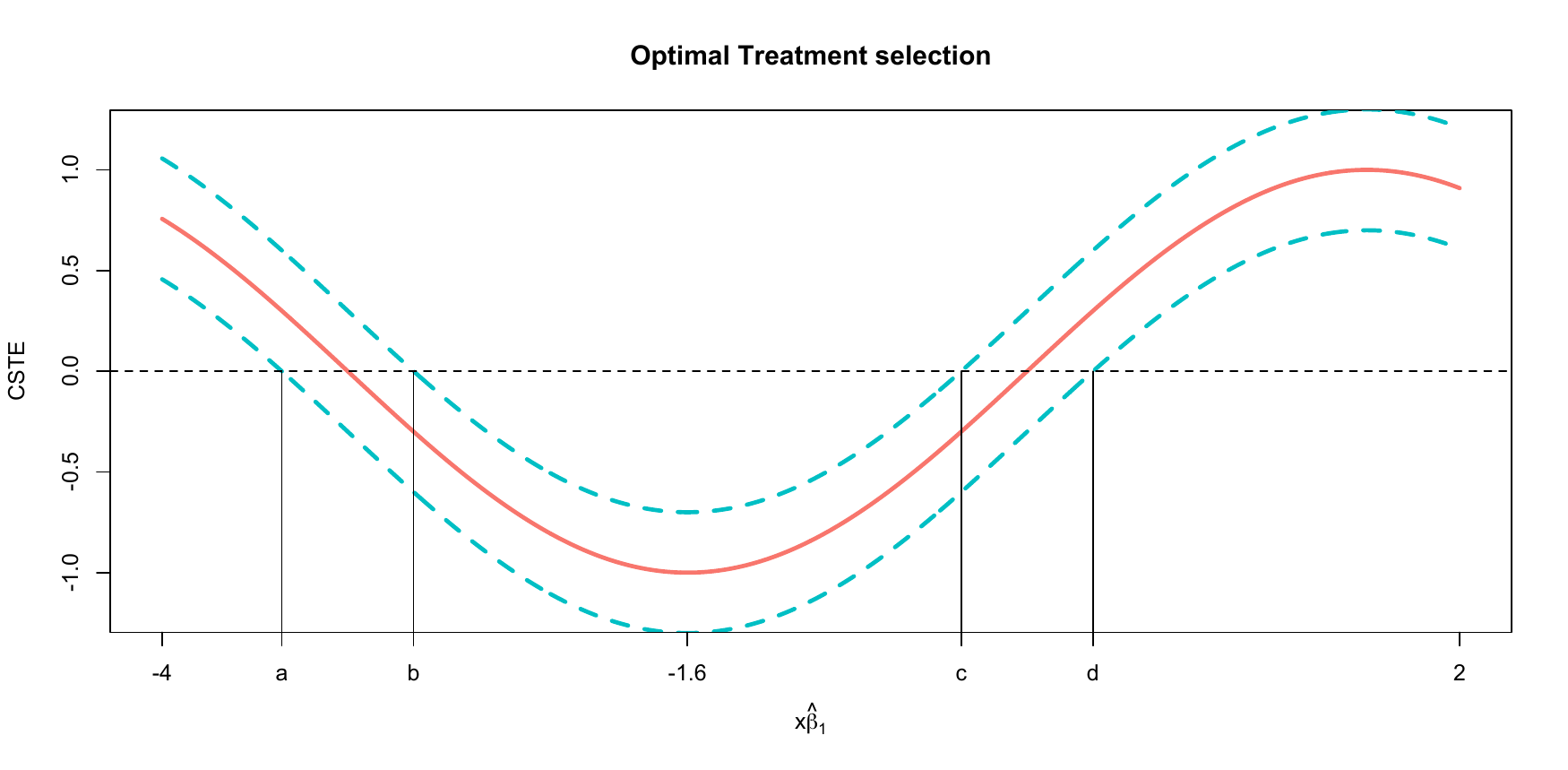}
\caption{A simulated example, demonstrating CSTE estimation, SCBs and cutoff points. The solid curve is the estimate of the CSTE
function. The dashed curves are the corresponding SCBs of the CSTE
curve. Four vertical lines indicate the locations of the cutoff points.}
\label{example}
\end{figure}

In this section, attention is focused on making individualized treatment
decision rule for patients. We provide an example to illustrate how to
select the optimal treatment based on the CSTE curve and its SCBs.
The goal of treatment selection is to find which group of patients will
benefit from new treatment based on their covariates. By the definition of
CSTE curve, if we assume that the outcome of interest is death, the CSTE
curve is the odds ratio of the treatment effect in reducing the probability
of death. That is, a positive ${\mathrm{CSTE}}(X)$ value means that the
patients will not benefit from new treatment since they may have a higher
death rate than patients who receive old treatment. We define the cutoff
points as the places where the upper and lower confidence bands equal to
0.
Based on these cutoff points, we are able to identify the regions with the
positive and negative values of ${\mathrm{CSTE}}(X)$, respectively, so that
it will guide us to select the best treatment for a future patient. To
summarize, this treatment selection method consists of the following steps:
\begin{step}
Use the existing dataset to obtain the SBK estimator $\widehat{g}%
_{1,sbk}(x^{\top }\widehat{\beta }_{1})$ of the CSTE curve and the corresponding SCBs  $\widehat{g}%
_{1,sbk}(x^{\top }\widehat{\beta }_{1})\pm \sigma _{n}(x^{\top }\widehat{%
\beta }_{1})Q_{h}(\alpha )$ at a given confidence level $100(1-\alpha)\%$.
\end{step}
\begin{step}
Identify the cutoff points for $x^{\top }\widehat{\beta }_{1}$ and the regions of positive and negative values of the constructed SCBs, as illustrated in Figure \ref{example}. 
\end{step}
\begin{step}
 For a new patient, we first calculate $\tilde{x}^{\top }\widehat{\beta }_{1}$, where $\tilde{x}$ is the observed value of the covariates for this new patient.  Then we make treatment recommendation for this patient based on what region the value of $x^{\top }\widehat{\beta }$ falls into. 

\end{step}

We use the following example to illustrate the method of using the
SCBs to select optimal treatment for patients. We
assume that $X^{\top }\hat{\beta}_{1}$ has a range of $(-4,2)$. In Figure %
\ref{example}, the solid line represents a CSTE curve and dashed lines above
and below the curve are the corresponding 95\% SCBs. We assume
that the outcome variable $Y$ is the indicator of death. As shown in Figure %
\ref{example}, the CSTE is decreasing when the value of $X^{\top }\hat{\beta}%
_{1}$ is from -4 to -1.6 and it is increasing when the value of $X^{\top }%
\hat{\beta}_{1}$ is from -1.6 to 2. In general, when the $X^{\top }\hat{\beta%
}_{1}$ value of a patient is within the range of (-4,-1.6), a larger value
implies that the patient more likely benefits from the new treatment than
from the old treatment. On the other hand, if the patient's $X^{\top }\hat{%
\beta}_{1}$ value falls into (-1.6,2), the new treatment is more beneficial
when a smaller value of $X^{\top }\hat{\beta}_{1}$ is observed. Moreover, the results in
Figure \ref{example} also imply that we are 95\% confident that a new
patient with the $X^{\top }\hat{\beta}_{1}$ value falling into the region $%
[-4,a]$ or $[d,2]$ will benefit from the old treatment, since the lower
bands are all above zero in this region. If $X_{0}^{\top }\hat{\beta}_{1}$
falls into region $[b,c]$, then we have 95\% confidence that the patient
should receive the new treatment, as the upper bands are all below zero in
this region. For the intervals $[a,b]$ and $[c,d]$, zero is covered in the
confidence band indicating that there is no significant difference between
the effects of the new and old treatments. 

It is worth nothing that the regions of indifference can also be constructed based on a parametric model using the limiting distribution of the parametric estimators; see more discussions in \cite{wu2016set}. However, this approach is only directly applicable to the settings with low-dimensional covariates and to each given patient instead of a group of patients. \citet{CSTE} define a modified
version of the CSTE curve using the quantile of the covariate to compare
covariate's capacities for predicting responses to a treatment. Then they
use the ``best" covariate as a guidance to select treatment. This may be
time-consuming when there is a large number of covariates. Our method
selects relevant covariates automatically in the estimation procedure and
combines information from all covariates through a weighted combination of
those covariates with the weights estimated from the data. The weight
reflects how important the corresponding covariate is for predicting the
response. As a result, our method is more convenient and flexible in
selecting optimal treatment.

\section{Simulation Study}
\label{simulation}
In this section, we investigate the finite-sample performance of our proposed method via simulated datasets. We run all simulations in \verb|R| in a linux cluster. We consider the following examples:

\begin{example}
$\logit(\mu(X,Z))=X^{\top}\beta_1(1-X^{\top}\beta_1)Z+\exp(X^{\top}\beta_2).$
\end{example}

\begin{example}
$\logit(\mu(X,Z))=(X^{\top}\beta_1)^2Z+\sin(\pi X^{\top}\beta_2/2).$
\end{example}

\begin{example}
$\logit(\mu(X,Z))= - \exp(X^{\top}\beta_1)Z/1.5 + (X^{\top}\beta_2)^2.$
\end{example}

\begin{example}
$\logit(\mu(X,Z))=X^{\top}\beta_1(1-X^{\top}\beta_1)Z+\exp(X^{\top}\beta_2)+c \cdot\sin(\pi X^{\top}\beta_3/2).$
\end{example}

\begin{example}
$\logit(\mu(X,Z))=[X^{\top}\beta_1(1-X^{\top}\beta_1) +c \cdot \sin(\pi X^{\top}\beta_3/2)] Z+\exp(X^{\top}\beta_2)$
\end{example}

The simulated data are generated as follows: the outcome $Y$ is sampled from a binomial distribution with probability of success equals to $\mu(x,z)$;  the covariates $X$ are generated from a truncated multivariate normal distribution with mean vector 0, covariance matrix with $\Sigma_{ij}=0.5^{|i-j|}$, and each covariate is truncated by $(-2, 2)$; the binary scale covariate $Z$ is sampled from $Binomial(1, 0.5)$, which means that each subject is randomly assigned to either control or treatment group. We set $\beta_1=(1,1,1,0, \dots, 0)'/\sqrt{3}$, $\beta_2=(1,-2,0,\dots,0)'/\sqrt{5}$, and $\beta_3=(1,1,1,0, \dots, 0)'/\sqrt{3}$. For examples 1-3, we set $p = 10, 50, 100, 500$, and sample size $n = 500, 750, 1000$. For examples 4-5, we consider consider that $g_2$ and $g_1$ are mis-specified respectively. We choose $p=50$, $n=1000$, and $c=0.1, 0.2, 0.5$. For each pair of $n$ and $p$, we repeat the simulations $J=300$ times. 

To obtain sparse estimates of $\beta_k$ for high-dimensional cases, we choose SCAD as the penalty function and let $\gamma=3.7$ \citep{MR1946581}. The optimal tuning parameter $\lambda$ is chosen from a geometrically increasing sequence of 30 parameters by minimizing the modified Bayesian information criterion \citep{wang2009shrinkage}:
$
BIC_{\lambda} = -2\cdot \log(Loss)+ df_{\lambda}\cdot \log(n) \cdot C(p),
$
where $Loss$ is the loss function in (\ref{EQ:Ln:sparse}), $df_{\lambda}$ is the number of non-zero elements in $\beta_k$, $C(p)=C(p)=\log\log(p)$.

We evaluate the following metrics for the non-parametric function $g_1(\cdot)$ based on 200 equally spaced grid points $t_1, \dots, t_{200}$ on the range of $\hat\eta_1=X^{\top }\hat{\beta}_{1}$ : the mean squared error (MSE) of its estimator; the mean absolute error (MAE) of its estimator; the average coverage probability (CP) of its SCBs. Since the models in Example 1-3 are correcly specified, we compute the average number of parameters that are incorrectly estimated to be non-zero, the average number of parameters that are incorrectly estimated as zero; the proportions that all relevant covariates are correctly selected and the proportions that some relevant covariates are not selected.

We define that the oracle estimator of $g_1(\cdot)$ is obtained when the true indexes of non-zero components in $\beta_1$ and $\beta_2$ are given in example 1-3. The results are summarized in Table \ref{sim-res} and 
\ref{sim-res-2}. In Table \ref{sim-res}, we see that the coverage probabilities are slightly less than 95\% but close to 90\% when the sample size is 750. When the sample is 1000, the empirical coverage is close to the nominal 95\% confidence level. The mean square error and mean absolute error also decrease as sample size $n$ increases. This indicates that the estimates of confidence bands become more accurate as the sample size increases. Table \ref{sim-res} shows that the proposed asymptotic SCB performs well when the sample size is relatively large. However, with small sample size, a bootstrap procedure is recommended to construct the SCB.  This can be a future research topic to explore. In Table \ref{sim-res-2}, when $c$ is small (i.e., the degree of misspecification is small), the numerical results are comparable to those for $c=0$ with no misspecification. However, the performance deteriorates when $c$ becomes large. Moreover, the method is more robust to the misspecification of $g_2$ than that of  $g_1$. The model selection results are summarized in Table \ref{rate}. The false positive and false negative are decreasing when $n$ increases. It is worth noting that in practice we often need to make a reliable treatment recommendation strategy to a group of patients instead of a specific given patient, and the SCBs can serve the former purpose well. In Table \ref{CI_SCB}, we also compare the performance of the pointwise confidence interval (CI) with that of the SCB for a group of randomly generated new patients with group size 1, 10, 100, 200, 500, respectively.  
The CI and SCB are constructed based on the data with sample $n=1000, p=10$. We see that the pointwise CI has low empirical coverage rates when the group size is greater than 1. The pointwise CI can work well when we only have one patient. However, when we have a group of patients with different covariate values,  we can make an incorrect treatment decision based on the pointwise CI for these patients.

\begin{table}
\caption{\label{sim-res}Simulation results for example 1-3. Oracle estimator is obtained when true index sets are given. True functions are $x(1-x), x^2, -\exp(x)/1.5$. MSE and MAE represent the mean squared error and the mean absolute error of the estimator for $g_1(\cdot)$, and CP represents the coverage probability of its SCBs.}
\centering
\resizebox{0.9\textwidth}{!}{%
\begin{tabular}{clccclccclccc}
               &  & \multicolumn{3}{c}{Model 1} &  & \multicolumn{3}{c}{Model 2} &  & \multicolumn{3}{c}{Model 3} \\
$(n, p)$       &  & MSE     & MAE     & CP      &  & MSE     & MAE     & CP      &  & MSE     & MAE     & CP      \\
(500, Oracle)  &  & 0.230   & 0.364   & 0.880   &  & 0.217   & 0.336   & 0.861   &  & 0.389   & 0.400   & 0.840   \\
(750, Oracle)  &  & 0.125   & 0.276   & 0.925   &  & 0.113   & 0.251   & 0.905   &  & 0.143   & 0.268   & 0.905   \\
(1000, Oracle) &  & 0.095   & 0.23    & 0.947   &  & 0.088   & 0.230   & 0.935   &  & 0.083   & 0.220   & 0.940   \\
(500, 10)      &  & 0.280   & 0.494   & 0.874   &  & 0.325   & 0.428   & 0.889   &  & 0.405   & 0.461   & 0.858   \\
(750, 10)      &  & 0.169   & 0.318   & 0.897   &  & 0.151   & 0.293   & 0.924   &  & 0.329   & 0.472   & 0.885   \\
(1000, 10)     &  & 0.112   & 0.258   & 0.939   &  & 0.095   & 0.243   & 0.929   &  & 0.099   & 0.241   & 0.931   \\
(500, 50)      &  & 0.530   & 0.461   & 0.871   &  & 0.506   & 0.576   & 0.870   &  & 0.679   & 0.621   & 0.863   \\
(750, 50)      &  & 0.258   & 0.371   & 0.930   &  & 0.331   & 0.426   & 0.903   &  & 0.280   & 0.393   & 0.920   \\
(1000, 50)     &  & 0.100   & 0.248   & 0.957   &  & 0.129   & 0.280   & 0.955   &  & 0.099   & 0.242   & 0.959   \\
(500, 100)     &  & 0.698   & 0.594   & 0.836   &  & 0.525   & 0.566   & 0.826   &  & 0.640   & 0.660   & 0.862   \\
(750, 100)     &  & 0.153   & 0.305   & 0.906   &  & 0.452   & 0.477   & 0.896   &  & 0.141   & 0.291   & 0.907   \\
(1000, 100)    &  & 0.108   & 0.230   & 0.947   &  & 0.188   & 0.338   & 0.920   &  & 0.093   & 0.237   & 0.940   \\
(500, 500)     &  & 0.718   & 0.789   & 0.826   &  & 0.642   & 0.777   & 0.835   &  & 0.683   & 0.519   & 0.831   \\
(750, 500)     &  & 0.413   & 0.503   & 0.857   &  & 0.434   & 0.453   & 0.860   &  & 0.524   & 0.444   & 0.854   \\
(1000, 500)    &  & 0.123   & 0.249   & 0.930   &  & 0.174   & 0.319   & 0.925   &  & 0.143   & 0.287   & 0.912  
\end{tabular}%
}
\end{table}

\begin{table}
\caption{\label{sim-res-2}Simulation results for example 4-5 (mis-specified cases) with $p=50$ and $n=1000$. MSE and MAE represent the mean squared error and the mean absolute error of the estimator for $g_1(\cdot)$, and CP represents the coverage probability of its SCBs.}
\centering
\label{compare-ci}
\begin{tabular}{cccccccc}
      & \multicolumn{3}{c}{Model 4} &  & \multicolumn{3}{c}{Model 5} \\
      & MSE      & MAE     & CP     &  & MSE      & MAE     & CP     \\
c = 0 & 0.100    & 0.248   & 0.957  &  & 0.100    & 0.248   & 0.957  \\
c=0.1 & 0.138    & 0.212   & 0.933  &  & 0.161    & 0.293   & 0.938  \\
c=0.2 & 0.136    & 0.235   & 0.931  &  & 0.266    & 0.338   & 0.914  \\
c=0.5 & 0.208    & 0.341   & 0.904  &  & 0.290    & 0.415   & 0.895 
\end{tabular}
\end{table}

\begin{table}
\caption{Comparisons between the pointwise confidence interval and the SCBs for the empirical coverage rate of a group of randomly generated new patients with group size 1, 10, 100, 200, and 500, respectively.  }
\centering
\label{CI_SCB}
\begin{tabular}{lllllll}
                               &  group size & 1     & 10  & 100     & 200   & 500  \\
\multirow{2}{*}{Model 1} &CI    & 0.962 & 0.806 & 0.621 & 0.561 & 0.554 \\
                         & SCBs & 1.000 & 0.991  & 0.952 & 0.948 & 0.935 \\ 

\multirow{2}{*}{Model 2} &CI    & 0.959 & 0.911 & 0.835 & 0.771 & 0.773 \\
                         & SCBs & 1.000 & 1.000  & 0.963 & 0.951 & 0.954 \\ 
\multirow{2}{*}{Model 3} & CI   & 0.961 & 0.853 & 0.694 & 0.661  & 0.668 \\
                         & SCBs & 1.000 & 0.981  & 0.952 & 0.934 & 0.937 
\end{tabular}
\end{table}

\begin{table}
\caption{Model selection results for example 1-3. FP(false positive): zero is estimated as non-zero; FN(false negative): non-zero is estimated as zero; C(correctly selected): all relevant covariates are selected; IC(incorrectly selected): some relevant covariates are not selected.}
\label{rate}
\centering
\resizebox{0.9\textwidth}{!}{%
\begin{tabular}{clcccllcccllcccl}
            &  & \multicolumn{4}{c}{Model 1}                    &  & \multicolumn{4}{c}{Model 2}                    &  & \multicolumn{4}{c}{Model 3}                    \\
$(n, p)$    &  & FPR   & FNR   & C     & \multicolumn{1}{c}{IC} &  & FPR   & FNR   & C     & \multicolumn{1}{c}{IC} &  & FPR   & FNR   & C     & \multicolumn{1}{c}{IC} \\
(500, 10)   &  & 0.282 & 0.025 & 0.885 & 0.115                  &  & 0.274 & 0.019 & 0.904 & 0.095                  &  & 0.381 & 0.015 & 0.909 & 0.090                  \\
(750, 10)   &  & 0.206 & 0.008 & 0.956 & 0.043                  &  & 0.170 & 0.007 & 0.960 & 0.039                  &  & 0.280 & 0.018 & 0.941 & 0.058                  \\
(1000, 10)  &  & 0.162 & 0.000 & 1.000 & 0.000                  &  & 0.110 & 0.005 & 0.994 & 0.005                  &  & 0.217 & 0.002 & 0.989 & 0.010                  \\
(500, 50)   &  & 0.271 & 0.042 & 0.785 & 0.214                  &  & 0.172 & 0.181 & 0.564 & 0.435                  &  & 0.165 & 0.236 & 0.535 & 0.465                  \\
(750, 50)   &  & 0.108 & 0.041 & 0.823 & 0.176                  &  & 0.178 & 0.025 & 0.892 & 0.107                  &  & 0.094 & 0.125 & 0.563 & 0.436                  \\
(1000, 50)  &  & 0.070 & 0.009 & 0.952 & 0.047                  &  & 0.111 & 0.005 & 0.975 & 0.025                  &  & 0.082 & 0.065 & 0.788 & 0.211                  \\
(500, 100)  &  & 0.063 & 0.093 & 0.625 & 0.375                  &  & 0.107 & 0.084 & 0.605 & 0.395                  &  & 0.093 & 0.087 & 0.670 & 0.330                  \\
(750, 100)  &  & 0.055 & 0.065 & 0.703 & 0.296                  &  & 0.098 & 0.044 & 0.688 & 0.311                  &  & 0.040 & 0.106 & 0.681 & 0.318                  \\
(1000, 100) &  & 0.031 & 0.032 & 0.845 & 0.154                  &  & 0.053 & 0.037 & 0.830 & 0.170                  &  & 0.021 & 0.061 & 0.803 & 0.196                  \\
(500, 500)  &  & 0.114 & 0.138 & 0.471 & 0.519                  &  & 0.119 & 0.164 & 0.223 & 0.777                  &  & 0.104 & 0.166 & 0.243 & 0.757                  \\
(750, 500)  &  & 0.974 & 0.096 & 0.540 & 0.460                  &  & 0.069 & 0.138 & 0.294 & 0.706                  &  & 0.059 & 0.132 & 0.302 & 0.698                  \\
(1000, 500) &  & 0.030 & 0.062 & 0.609 & 0.391                  &  & 0.039 & 0.102 & 0.413 & 0.587                  &  & 0.049 & 0.094 & 0.449 & 0.551                 
\end{tabular}%
}
\end{table}

\section{A Real Data Example}
\label{real-world}
In this section, we present and discuss the results of applying the procedure described in previous sections to a real data set. We illustrate the applications of the CSTE curve in a real-world example and provide some insight into the interpretation of the results. The goal is to analyze the effect of Zhengtianwan in the treatment of migraines. This is a multicenter, randomized, double-blind, placebo-controlled trial on the effectiveness of Zhengtian pill on treating patients with migraines. A migraine is a common neurological disorder characterized by recurrent headache attacks. Migraine treatment involves acute and prophylactic therapy. The objective of this study is to evaluate the efficacy and safety of Zhengtian Pill for migraine prophylaxis. Zhengtian Pill, a Chinese Patented Medicine approved by the State Food and Drug Administration of China in 1987, has been used in clinical practice for more than 20 years in China to stimulate blood circulation, dredge collaterals, alleviate pains, and nourish the liver. To evaluate the effectiveness of Zhengtian Pill in preventing the onset of migraine attacks, a large-scale, randomized and prospective clinical study was conducted. Eligible patients were monitored during a baseline period of four weeks, during which the headache characteristics were recorded as baseline data. In this period, any use of migraine preventive medications was prohibited. After the baseline period, a 12-week treatment period and four-week follow-up period were carried out. Patients were requested to keep a headache diary throughout the whole study period, from which investigators were able to extract detailed information of migraine attacks including migraine days, frequency, duration, and intensity as well as the use of acute medication during the study period. The outcome measures were evaluated at 4, 8, and 12 weeks, and during the follow-up period. The patients who met the inclusion criteria were randomly assigned into the experimental group and control group in a 1:1 ratio using a computer-generated stochastic system.

In our analysis, the response variable $Y$ is a binary outcome indicating if the number of days that headaches occur has decreased 8 weeks after patients were treated. $Z$ is another indicator: $Z=1$ means the subject is assigned in the experimental group; $Z=0$ means in the control group. The covariates $x_1$ to $x_3$ are gender (0 for male, 1 for female), height and body weight, $x_4$ to $x_{12}$ are overall scores of Traditional Chinese Medicine Syndromes (Huozheng, Fengzheng, Xueyu, Tanshi, Qixu, Yuzheng, Xuexu, Yinxu, and Yangxu) for migraine. Those scores, with a range of 0 to 20, are commonly used to describe the severity of migraine symptoms in Traditional Chinese Medicine. The higher of the score means that the syndrome is more severe. All covariates are centered and standardized as input. We have 204 observations where 99 are in the experimental group and 105 are in the control group. The purpose of this exercise is to model the odds ratio as a function of those covariates. The number of covariates $p=12$ might be regarded as small, so we estimate the CSTE curve using the algorithm in Section \ref{sparse-logit} with and without model selection. We use SCAD as our penalty function and the optimal tuning parameter is selected via the modified BIC criterion. The corresponding SCBs for the CSTE curve are calculated. The results are not intended as definitive analyses of these data.

\begin{figure}[ht]
\centering
\subfigure[Without variable selection]{\includegraphics[width=0.45\textwidth]{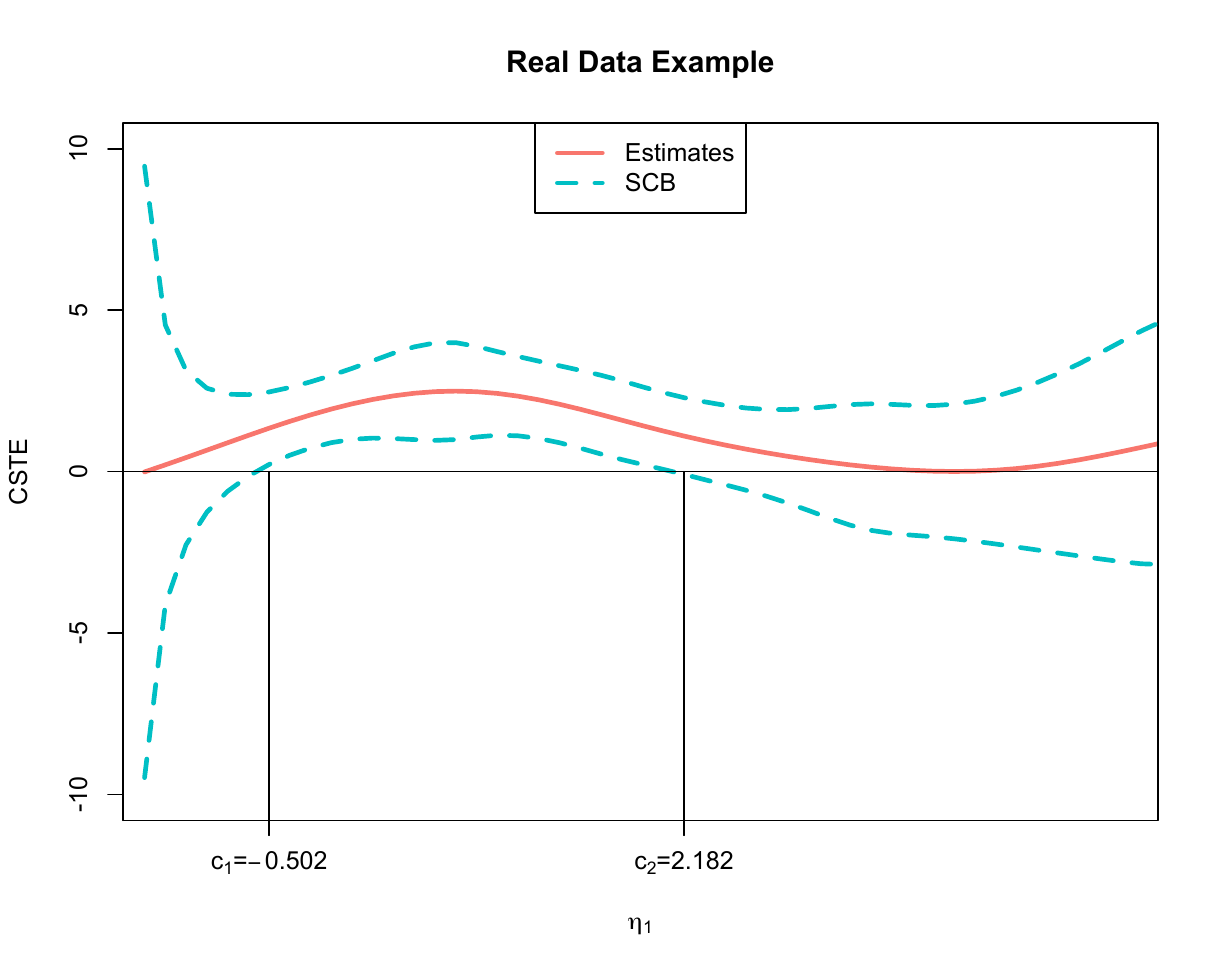}}
\subfigure[With variable selection]{\includegraphics[width=0.45\textwidth]{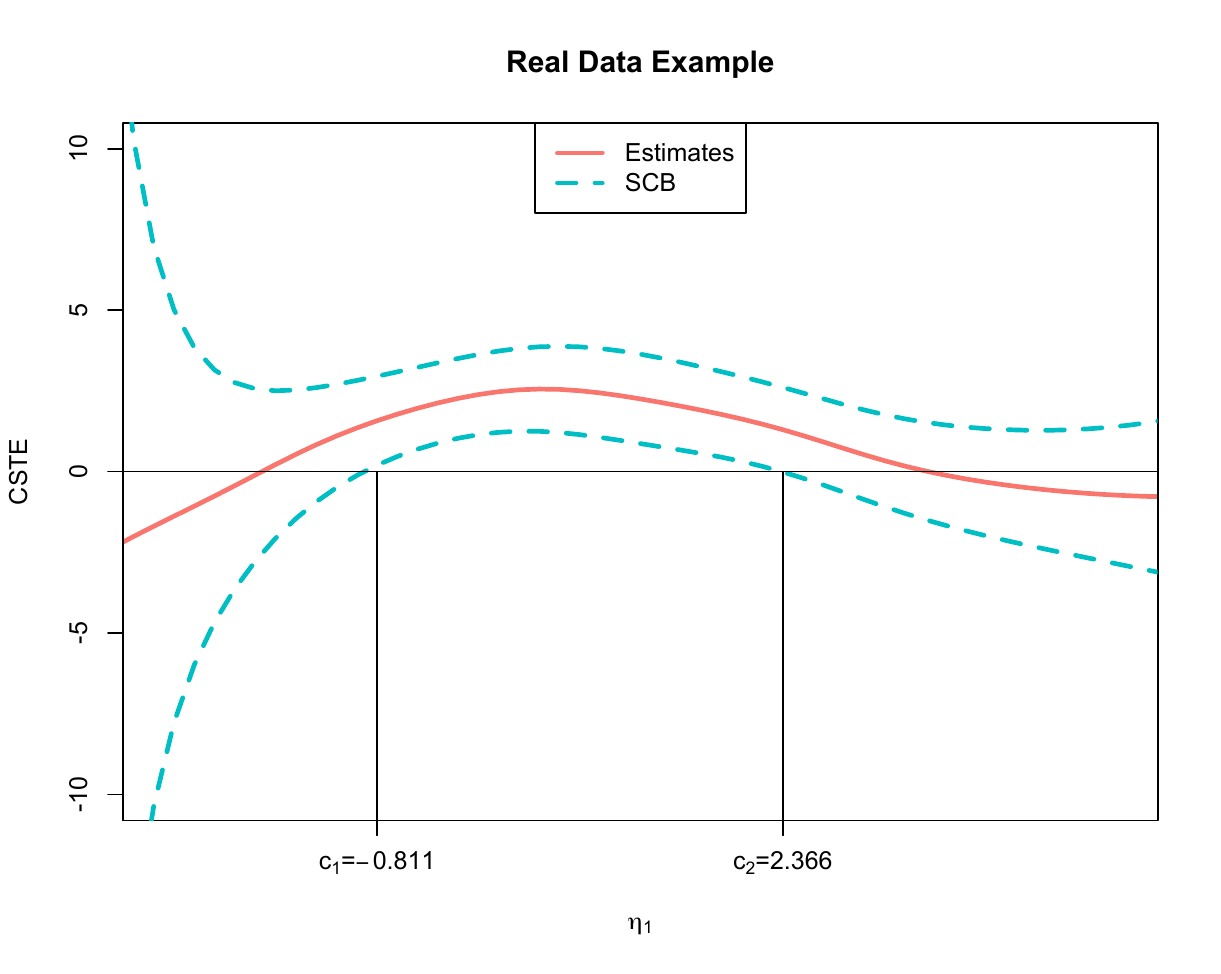}}
\caption{Real data example. Red curve is the CSTE curve; blue dashed curves are the SCBs. (a) two cutoff points are -0.502 and 2.182; (b) two cutoff points are -0.811 and 2.366.}
\label{real}
\end{figure}

\begin{figure}[ht]
\centering
\subfigure[Huozheng]{\includegraphics[width=0.32\textwidth]{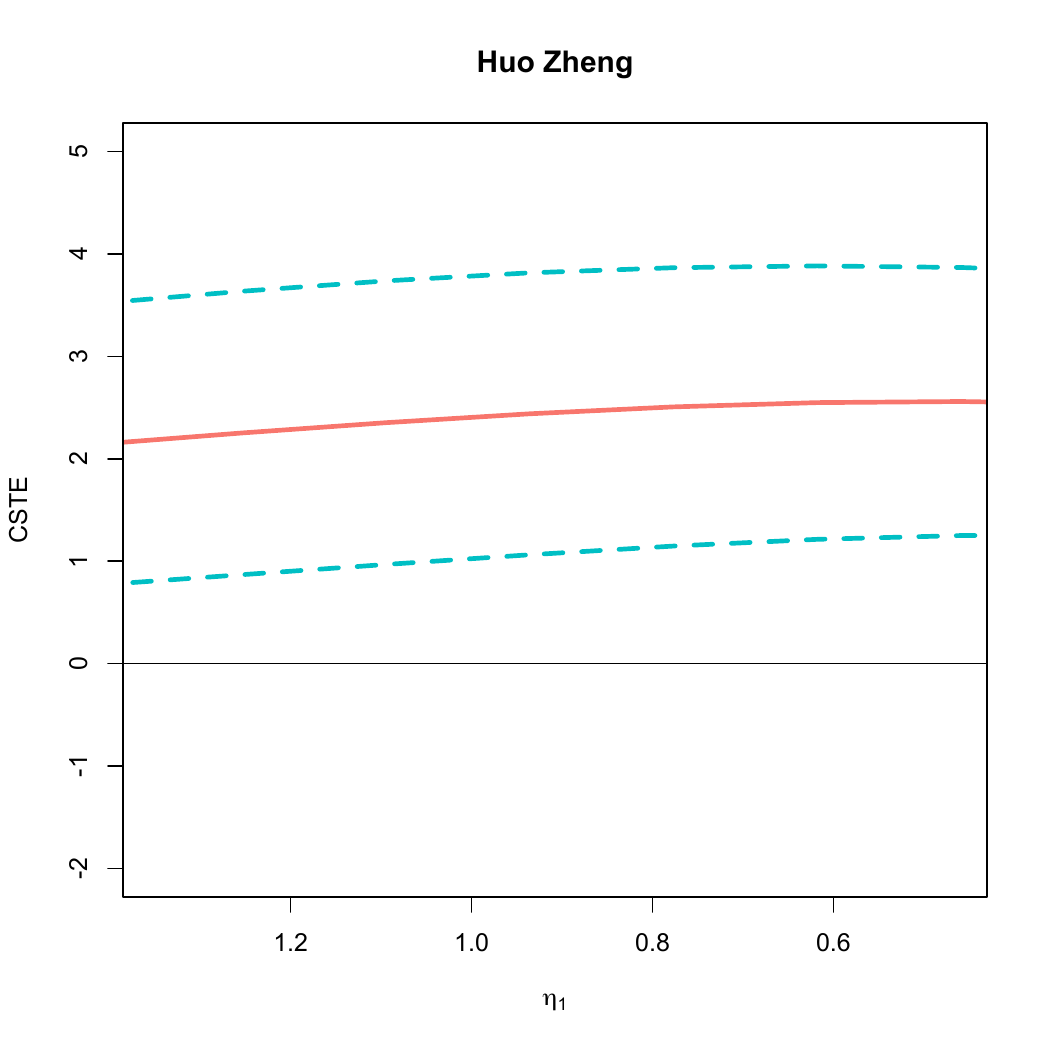}}
\subfigure[Fengzheng]{\includegraphics[width=0.32\textwidth]{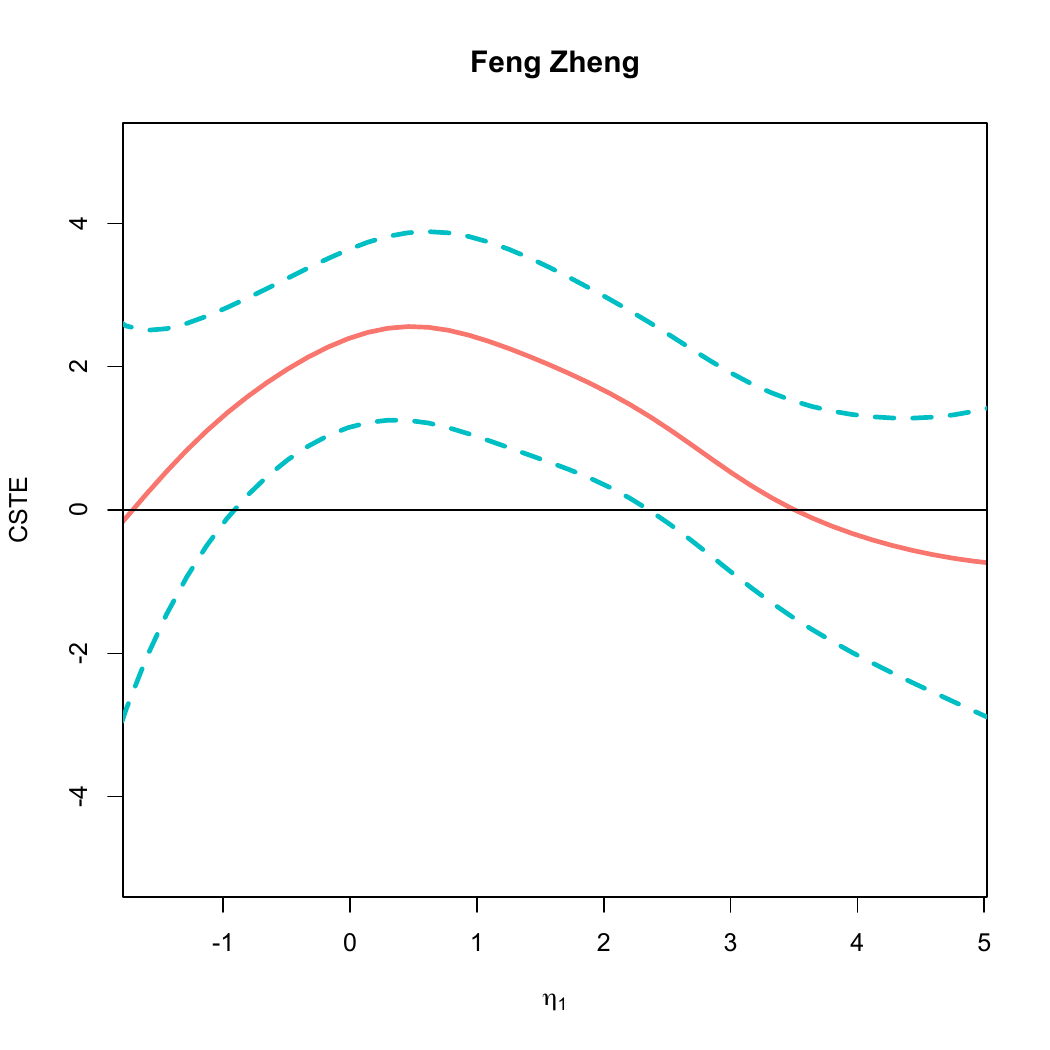}}
\subfigure[Xueyu]{\includegraphics[width=0.32\textwidth]{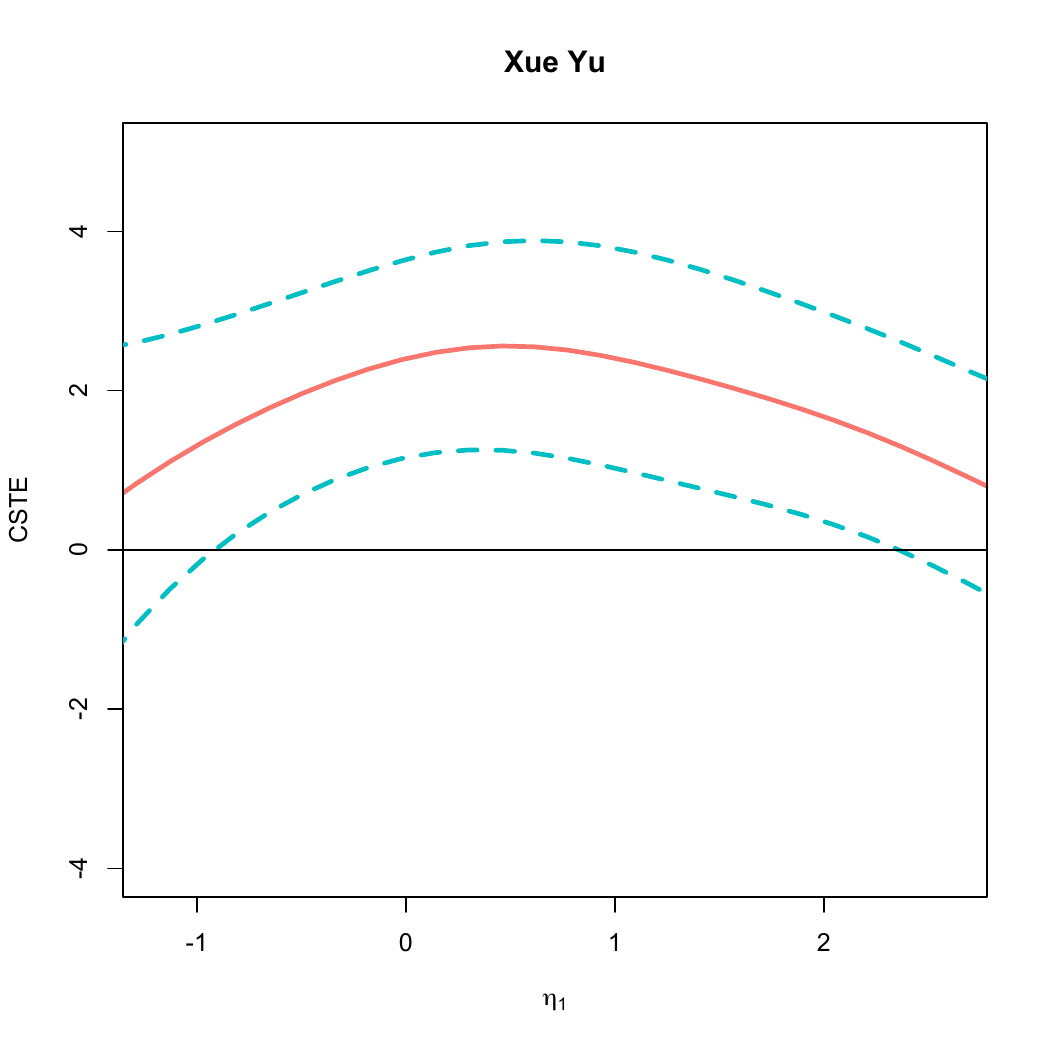}}
\subfigure[Yuzheng]{\includegraphics[width=0.32\textwidth]{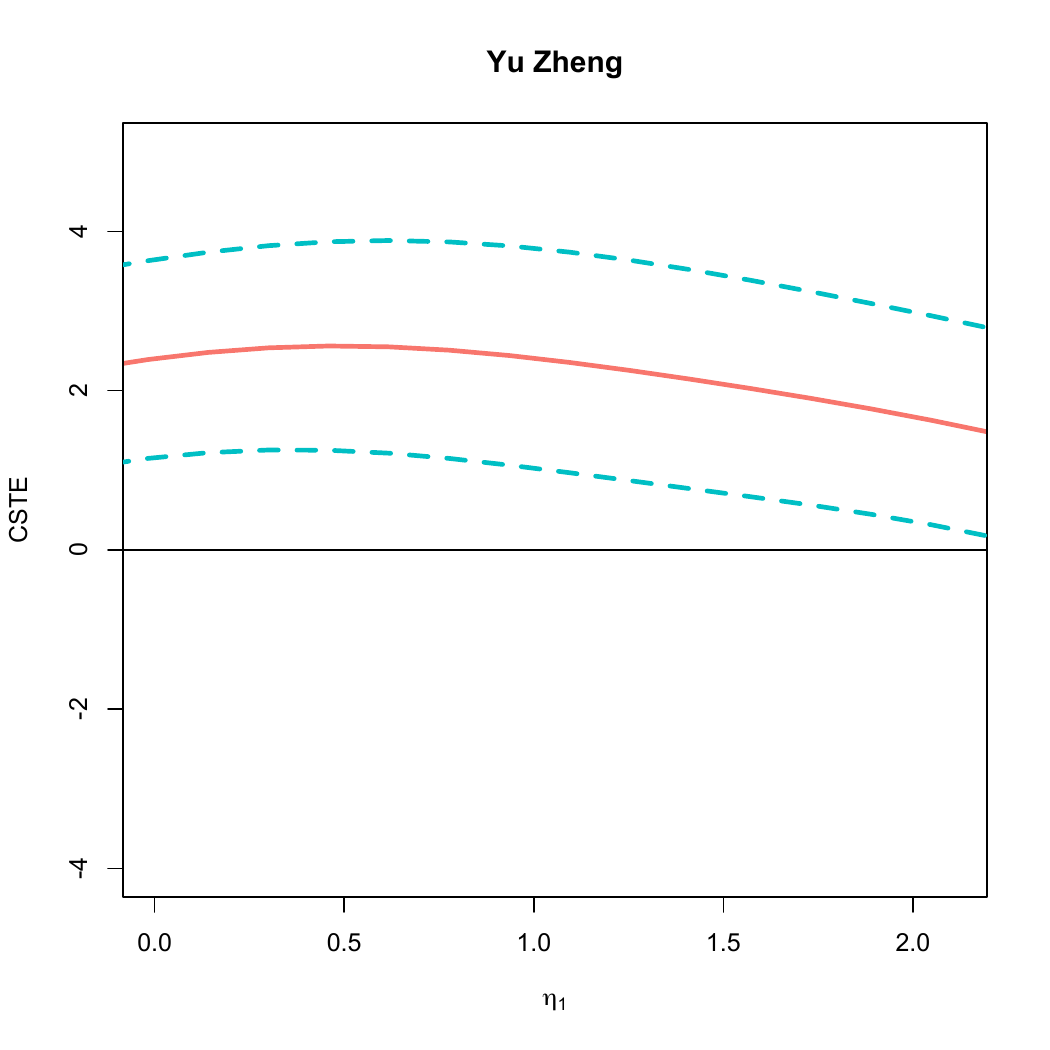}}
\subfigure[Xuexu]{\includegraphics[width=0.32\textwidth]{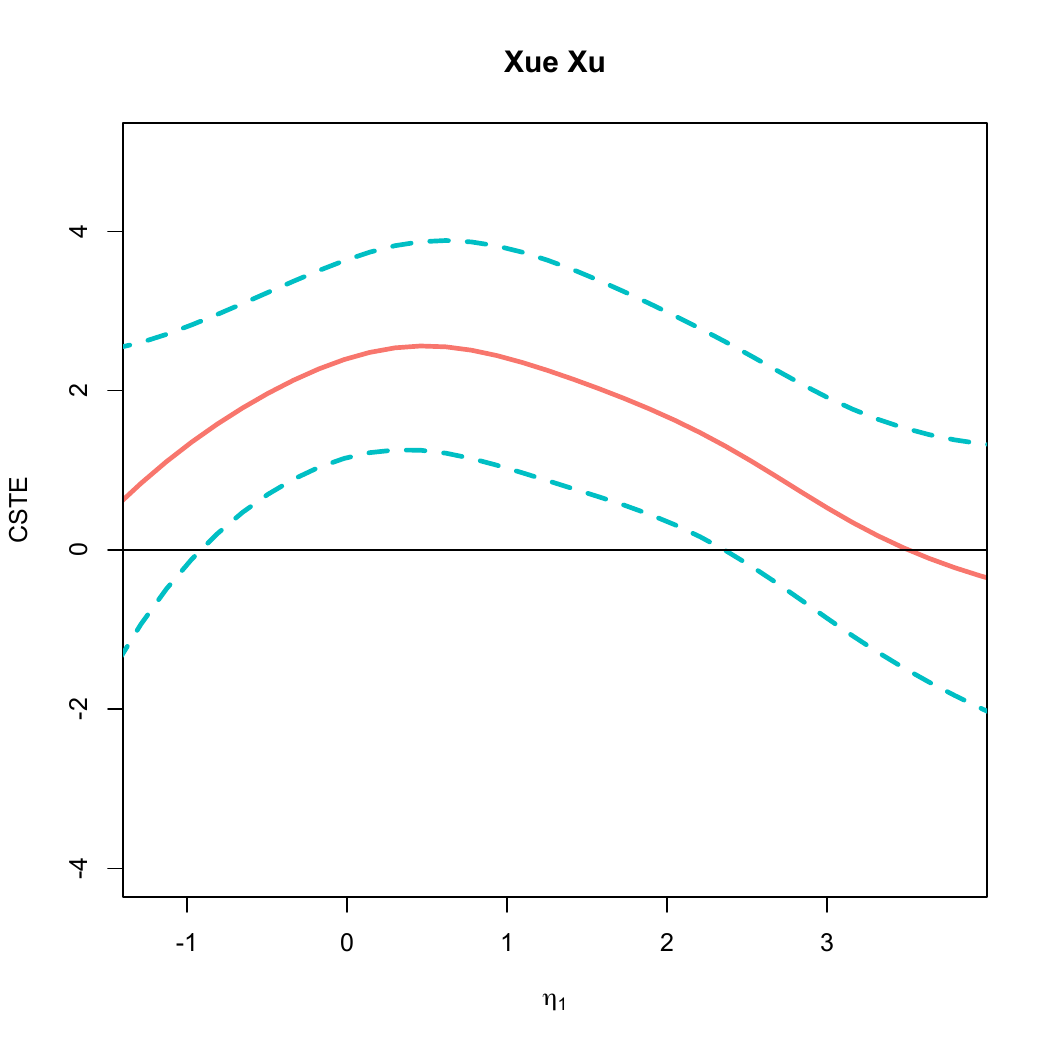}}
\subfigure[Yinxu]{\includegraphics[width=0.32\textwidth]{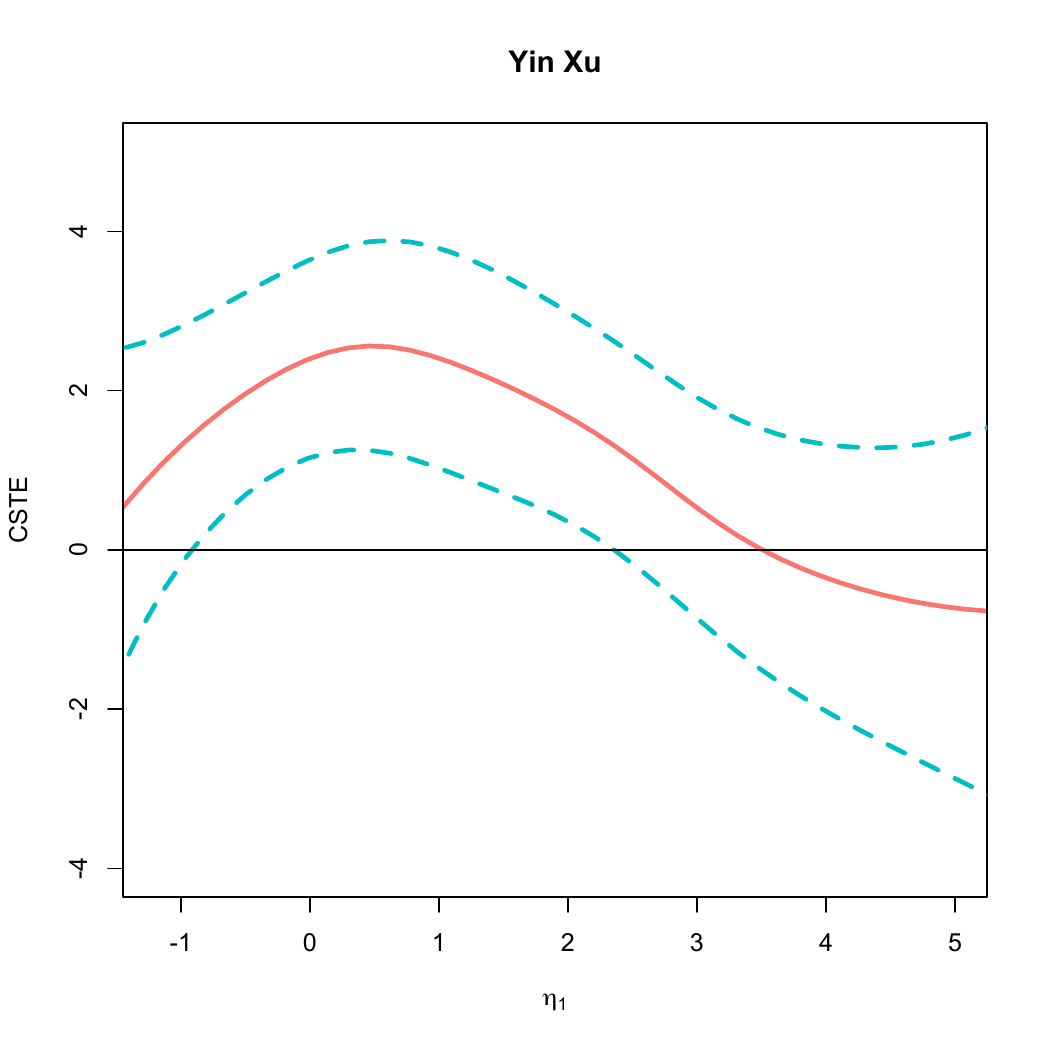}}
\subfigure[Yangxu]{\includegraphics[width=0.32\textwidth]{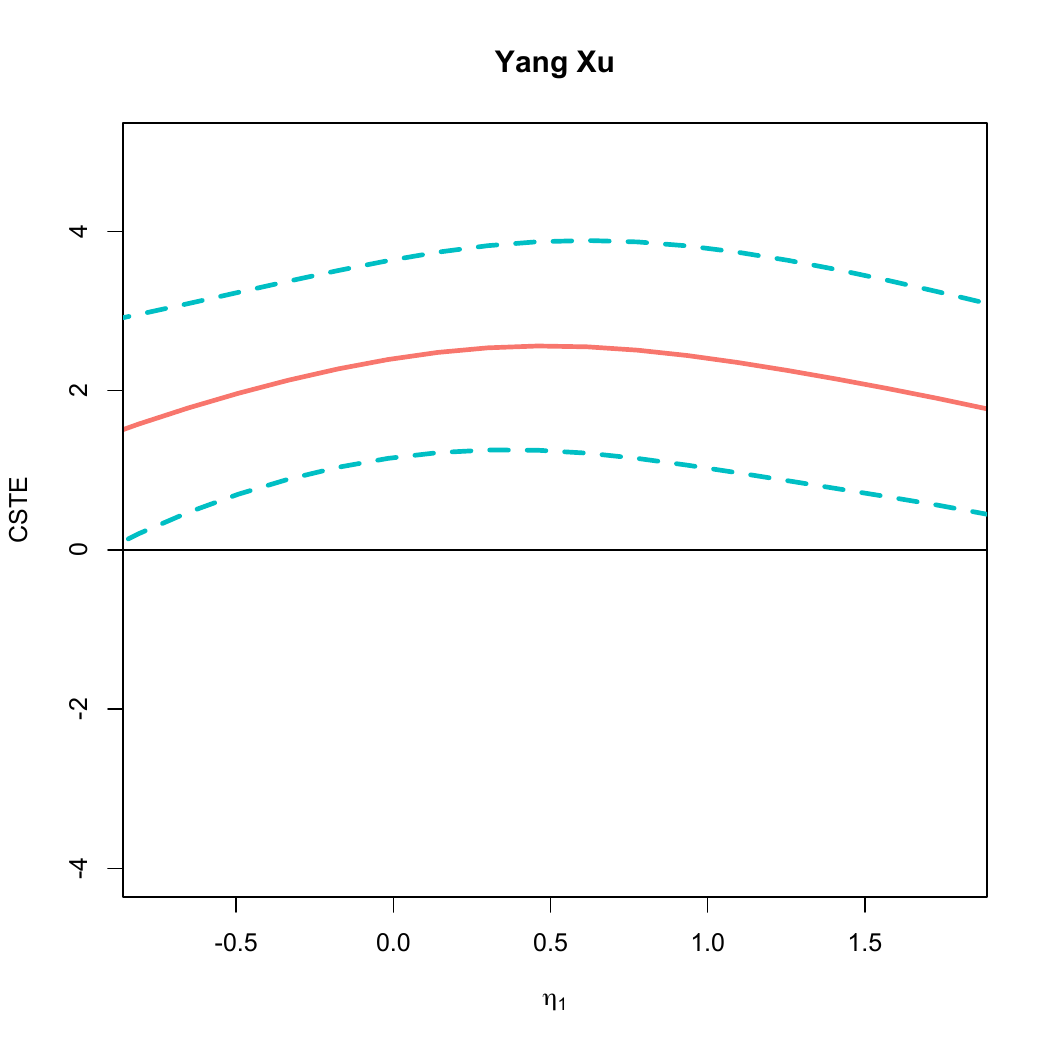}}
\caption{Plot one covariate against CSTE curve with other covariates fixed on their mean values.}
\label{real2}
\end{figure}

Table \ref{coef} summarizes the point estimates of $\beta_1$ and the corresponding standard errors. Figure \ref{real} shows two estimated CSTE curve and their SCBs: (a) using all 12 variables; (b) using all variables except $x_7$ and $x_8$ (not selected). To aid in interpretation for each covariate, we depict each covariate versus $g_1$ where other covariates are projected onto their mean values in Figure \ref{real2}. As we can see, Huozheng has a monotonic dependence on the CSTE, but the corresponding relationships for other overall scores are highly nonlinear; most of them have a quadratic appearance. On the basis of these Figures, we can conclude that the odds ratio does depend on the linear combination of the covariates in a nonlinear manner.

When all variables are used to estimate the curve, the two cutoff points are $c_1= -0.502, c_2=2.182$. The estimates of the overall rating of biomarker values are $x^{\top}\hat\beta_1$ where the majority of the points ($>$ 95\%) fall into the interval $(-1.2, 5)$. Two cutoff points divide this interval into three parts. Since the response variable $y=1$ represents headache improvements after 8 weeks treatment, higher CSTE value means that the patient more likely benefits from the treatment. Suppose a new patient with biomarker value $x_0$. When $x^{\top}\hat\beta_1$ falls into $[-1.2, c_1]$ and $[c_2, 5]$, the treatment does not have a significant effect. When it falls into $(c_1, c_2)$, then this treatment can improve the migraine of this patient. In the Zhengtianwan data, there are 37.7\% of patients fall into the interval. On the other hand, when $x_7, x_8$ are removed from the model, we obtain a new estimate $\hat\beta_{1,ms}$ of $\beta_{1}$. As we can see from Figure \ref{real} (b), the positive region of the CSTE curve becomes wider than that in (a), and the two cutoff points become -0.811 and 2.366. The majority of $x^{\top}\hat\beta_{1,ms}$ falls into $-(2.5, 5)$. When $x^{\top}\hat\beta_{1,ms}$ is in the range of $[-0.811, 2.366]$, the patient should receive Zhengtianwan treatment and 50.98\% of all subjects in the training dataset fall into this interval. 

The number of covariates is not very large in this dataset, so it is possible to fit a sparse linear model with high order terms and make treatment recommendation  based on the parametric estimate of the CSTE curve. To this end, we model $g_k$ for $k=1,2$ as a multivariate polynomial function of the covariates $X$ with degree of two, i.e. $g_k(X)=\alpha_{0, k}+\sum_{j} \alpha_{j, k} X_j+\sum_{j\le j'} \alpha_{jj', k} X_jX_{j'}.$ We obtain the estimates ($\hat\alpha_{0, k}$,$\hat\alpha_{j, k}$,$\hat\alpha_{jj', k}$) of the unknown coefficients ($\alpha_{0, k}$,$\alpha_{j, k}$,$\alpha_{jj', k}$) by fitting regularized $L_1$ logistic regression with the tuning parameter selected by 3-fold cross-validation. For a new patient with the observed covariates $\tilde{x}$, we obtain the predicted value for the CSTE curve as $\hat{g_1}(\tilde{x})=\hat\alpha_{0, 1}+\sum_{j} \hat\alpha_{j, 1} \tilde{x}_j+\sum_{j\le j'} \hat\alpha_{jj', 1} \tilde{x}_j\tilde{x}_{j'}$. For this parametric modeling method, we can only use the predicted value for $g_1$ to make treatment recommendation. Thus, based on the rule that a subject would benefit from the treatment if $\hat g_1$ is positive,  $67.64\%$ of the subjects should receive the treatment. We see that this percentage is higher than that obtained by our proposed method, as the parametric method only uses the point estimate for treatment selection. The construction of confidence intervals for $g_1$ by the parametric method is very challenging, as it involves finding the joint asymptotic distribution of ($\hat\alpha_{0, k}$,$\hat\alpha_{j, k}$,$\hat\alpha_{jj', k}$) after model selection, which is a difficult problem. Moreover, SCBs are not available based on the parametric modeling method. As discussed and shown in the simulation studies given in Section \ref{simulation}, SCBs are more reliable than the pointwise CIs for treatment recommendation for a group of patients.

\begin{table}
\caption{Estimates of $\beta_1$ and the corresponding standard errors (in parentheses). First row: using the unpenalized model; second row: using the penalized model.}
\label{coef}
\centering
\resizebox{1\textwidth}{!}{%
\begin{tabular}{ccccccccccccc}
Variable                               & Gender & Height & Weight & Huozheng & Fengzheng & Xueyu  & Tanshi & Qixu   & Yuzheng & Xuexu  & Yinxu  & Yangxu \\
\multirow{2}{*}{$\beta_1$}            & 0.88   & 0.10   & -0.04  & -0.06    & 0.08      & -0.09  & -0.08  & 0.04   & 0.18    & -0.24  & 0.30   & -0.03  \\
                                       & (0.23) & (0.01) & (0.01) & (0.03)   & (0.02)    & (0.01) & (0.03) & (0.11) & (0.04)  & (0.05) & (0.07) & (0.02) \\
\multirow{2}{*}{Penalizd $\beta_{1}$} & 0.70   & 0.11   & -0.05  & -0.04    & 0.14      & -0.15  & 0.00   & 0.00   & 0.21    & -0.38  & 0.49   & -0.14  \\
                                       & (0.31) & (0.02) & (0.01) & (0.02)   & (0.03)    & (0.01) & (0.00) & (0.00) & (0.05)  & (0.07) & (0.09) & (0.06) \\
\end{tabular}%
}
\end{table}

\section{Discussion}

\label{Discussion} Both the simulation and real-world studies in Sections \ref%
{simulation} and \ref{real-world} suggest that the modeling procedure for
CSTE curve can successfully detect and model complicated non-linear
relationships between binary response and high-dimensional covariates. In
practice, the non-linear dependencies we suggest are not characteristic of
all situations. We adapt the spline-backfitted kernel smoothing to construct
the SCBs for the non-linear functions to choose the
optimal treatment. Moreover, the SCBs can be used to verify the
presence of non-linear relationships as well.

Our model is motivated by the desire to provide an individualized decision
rule for patients along with the ability to deal with high-dimensional
covariates when the outcome is binary. The semi-parametric modeling approach can be
viewed as a generalization of the CSTE curve with one covariate proposed in \cite{CSTE}
in the sense that the odds ratio depends on a weighted linear combination of all
covariates. Although we consider a single decision with two treatment
options, our model can be readily generalized to multiple treatment arms.

To identify model (\ref{semi-model}), we require that the unknown functions $g_{1}\left( \cdot \right) $ and $g_{2}\left( \cdot \right) $ be
nonconstant functions on their supports. Otherwise, the index parameters $\beta_1$ and $\beta_2$ are not identified. That is, if $g_{k}\left( \beta
_{k}^{\top }x\right) =a_{k}$ for all $x\in {\cal X}$, then $\beta _{k}$ can
take an arbitrary value. Our treatment recommendation method is proposed
based on the prior information that at least one baseline
covariate is useful, and the treatment
effect is not zero for some values of the covariates. Based on this prior
information, our goal is to estimate the optimal treatment regimes and
provide recommendations for individual patients based on their observed
baseline covariates. For example,  Figures  \ref{example}-\ref{real} show that the estimated values of $g_{1}\left( \cdot
\right) $ are nonzero in some regions, so that one can apply our method to identify those regions and then make treatment recommendations to patients. In practice, if we conjecture that all covariates are irrelevant, then for an arbitrarily given value of $\beta _{k}$, $%
g_{k}\left( \beta _{k}^{\top }x\right) =a_{k}$ for all $x\in {\cal X}$. We can conduct a test for this conjecture by choosing a value for $\beta _{k}$, estimating the unknown function $%
g_{k}\left( \beta _{k}^{\top }x\right)$ and then testing whether the function is a constant or not. Likewise, we can use this approach to
testing whether $g_{1}\left( \beta _{1}^{\top }x\right) =0$ for all $x\in {\cal X}$, i.e. whether the
treatment has effect at all. If they are rejected, next we
can fit our model to more precisely find the optimal value of $\beta _{k}$. Since this paper
focuses on making individualized treatment recommendations to patients based on their available covariates,
we leave this interesting topic as a future work to explore.


\bigskip
\begin{center}
{\large\bf SUPPLEMENTARY MATERIAL}
\end{center}

\noindent Supplementary material includes the technical proofs for all the theoretical results.
\bibliographystyle{chicago}
\bibliography{ml}

\end{document}